\documentclass[twocolumn]{aastex63}
\pdfoutput=1

\newcommand{\cca}{Center for Computational Astrophysics, Flatiron Institute, Simons Foundation, 162 Fifth Avenue, New York, NY 10010, USA}

\usepackage{enumitem}
\usepackage{amsmath}
\usepackage{lineno}
\received{June 1, 2019}
\revised{January 10, 2019}
\accepted{\today}
\submitjournal{ApJ}

\shorttitle{Impact of LMC on Satellite Orbital Poles}
\shortauthors{Garavito-Camargo et al.}
\graphicspath{{./}{figures/}}
\begin{document}

%\title{The Milky Way's plane of satellites: a result of a non-inertial reference frame induced by the LMC}
%\title{Phase Space Bias in the Orbital Poles of Milky Way Satellites Induced by the LMC}
\title{The Clustering of Orbital Poles Induced by the LMC: Hints for the Origin of Planes of Satellites}

\author[0000-0001-7107-1744]{Nicol\'as Garavito-Camargo}
\affil{Steward Observatory, University of Arizona, 933 North Cherry Avenue,Tucson, AZ 85721, USA.}

\author[0000-0002-9820-1219]{Ekta Patel}
\affiliation{Department of Astronomy, University of California, Berkeley, 501 Campbell Hall, Berkeley, CA, 94720, USA} 
\affiliation{Miller Institute for Basic Research in Science, 468 Donner Lab,Berkeley, CA 94720, USA}

\author[0000-0003-0715-2173]{Gurtina Besla}
\affiliation{Steward Observatory, University of Arizona, 933 North Cherry Avenue,Tucson, AZ 85721, USA.}

\author[0000-0003-0872-7098]{Adrian~M.~Price-Whelan}
\affiliation{\cca}

\author{Facundo A. G\'omez}
\affiliation{Instituto de Investigaci\'on Multidisciplinar en Ciencia y Tecnolog\'ia, Universidad de La Serena, Ra\'ul Bitr\'an 1305, La Serena, Chile.}
\affiliation{Departamento de F\'isica y Astronom\'ia, Universidad de La Serena, Av. Juan Cisternas 1200 N, La Serena, Chile.}

\author[0000-0003-3922-7336]{Chervin F.P Laporte}
\affiliation{Institut de Ci\`{e}ncies del Cosmos (ICCUB), Universitat de Barcelona (IEEC-UB), Mart\'{i} i Franqu\`{e}s 1, 08028 Barcelona, Spain}
\affiliation{Kavli Institute for the Physics and Mathematics of the Universe (WPI), The University of Tokyo Institutes for Advanced Study (UTIAS), The University of Tokyo, Chiba 277-8583, Japan}

\author[0000-0001-6244-6727]{Kathryn V. Johnston}
\affiliation{Department of Astronomy, Columbia University, New York, NY 10027, USA.}
\affiliation{\cca}

\begin{abstract}
A significant fraction of Milky Way (MW) satellites exhibit phase-space properties consistent with a coherent orbital plane. Using tailored N--body simulations of
a spherical MW halo that recently captured a massive (1.8$\times 10^{11}$M$\odot$) LMC-like satellite, 
we identify the physical mechanisms that may enhance the clustering of orbital poles of objects orbiting the MW. 
The LMC deviates the orbital poles of MW dark matter (DM) particles from the present-day random distribution. Instead, the orbital poles of particles 
beyond $R\approx 50$kpc cluster near the present-day orbital pole of the LMC along a sinusoidal pattern across the sky. 
The density of orbital poles is enhanced near the LMC by a factor $\delta \rho_{max}$=30\%(50\%) with respect to underdense regions, and $\delta \rho_{iso}$=15\%(30\%) relative to the isolated MW simulation (no LMC) between 50-150 kpc (150-300 kpc).
The clustering appears after the LMC's pericenter ($\approx$ 50 Myr ago, 49 kpc) and lasts for at least 1 Gyr. 
Clustering occurs because of three effects: 1) the LMC shifts the velocity and position of the central density of the MW's halo and disk; 2) the DM dynamical friction wake and collective response  induced by the LMC 
changes the kinematics of particles;  3) observations of particles selected 
within spatial planes suffer from a bias, such that measuring orbital poles in a great circle in the sky enhances the probability of their orbital poles being clustered. This scenario should be ubiquitous in hosts that recently captured a massive satellite (at least $\approx$ 1:10 mass ratio), causing the clustering of orbital poles of halo tracers.
\end{abstract}

%% Keywords should appear after the \end{abstract} command. 
%% See the online documentation for the full list of available subject
%% keywords and the rules for their use.


%%%%%%%%%%%%%%%%%%%%%%
\section{Introduction} \label{sec:intro}

The idea that a number of satellite galaxies around the Milky Way (MW)
are arranged in a planar configuration has been discussed since the 1970s \citep[e.g.,][]{Kunkel1976, LyndenBell1976}. While this idea has been met with controversy,
 recent astrometric measurements \citep{Piatek03, Piatek05, Piatek07, Pryor15, Kallivayalil13, Sohn17, Fritz2018} 
have confirmed that more than half of the classical satellites of the MW and some of the more recently discovered ultra-faint dwarf galaxies \citep{Pawlowski2015, Fritz2018, Pawlowski2020, Li21} align within $\approx$37$^{\circ}$ of a common orbital pole and have a planar configuration of positions ($\approx$20--30 kpc root-mean-square height). Satellites in this plane, which is referred to as the Vast Polar Structure \citep[VPOS;][]{Pawlowski2012, Pawlowski2013, Pawlowski2015, Santos19, Santos20}, exhibit some amount of coherent co-rotation or counter-rotation. However, a physical explanation for the origin of the MW's VPOS remains yet unknown.

The existence of the VPOS has been challenging to explain in a cosmological context \citep[see][for a review]{Pawlowski2018}. As argued by \citet{Kroupa2005}, the observed distribution of MW satellites is not consistent with the distribution of cosmological subhalos about MW analogs, giving rise to the ``planes of satellites challenge" to the CDM model. Several subsequent studies have analyzed both large volume, cosmological simulations and higher resolution cosmological zoom-in simulations, finding promising links between planar associations of satellites and filamentary accretion or large-scale structure \citep[e.g.,][]{Zentner2005, Lovell2011, Libeskind2005, Libeskind2011, Libeskind2015}.
But the satellite planes formed through these mechanisms are generally thicker than the MW VPOS \citep{Pawlowski2020, Metz2009}.
However, an important caveat to this apparent challenge is that the VPOS is typically assumed (implicitly or explicitly) to be a long-lived structure rather than a transient feature.

One natural mechanism for generating correlated orbits is to accrete systems in a group \citep{Donghia08, Li2008, Metz2009, Sales17, Samuel20, santos20b}.

The challenge in reconciling the existence of the VPOS with cosmological models largely stems from the unknown physical origin of the alignment, making it difficult to connect the existence of a plane to properties of a given cosmological model. Recently, \citet{Samuel20} used a handful of systems from the FIRE suite of simulations to examine planes of satellites around MW-mass hosts and found that such planes may be more common around MW hosts that have recently captured an LMC analog. In this work we describe a physical mechanism that induces an apparent coherence in the phase-space properties of simulated dark matter (DM) particles in the MW's halo after the passage of the LMC. This coherence is a key signature of the observed VPOS structure. 

As described in \cite{Gomez15}, the orbital barycenter of the MW--LMC system shifts away from the center of the MW's disk as the LMC passes through pericenter. Consequently, an all-sky dipole pattern in radial velocities is expected, representing the reflex of the disk's motion \citep[see, e.g.][]{Gomez15,Erkal19, garavito-camargo19a, Peterson20, Cunningham20}. Excitingly, potential evidence of this ``reflex motion'' has been recently reported \citep{Erkal20c, Petersen20b}. In addition, \citep[][hereafter G20]{garavito-camargo2020} demonstrate that the center of mass of the outer halo ($>$ 30 kpc) is not expected to be coincident with the central density peak of the halo. Recently, \citet[][hereafter, G19]{garavito-camargo19a} and G20 demonstrate that the passage of the LMC will induce overdensities in the MW's DM halo, broadly characterized as a dynamical friction DM wake that trails the LMC, and a large-scale collective response that generally leads the LMC and is related to the COM displacement. 

The effect from the LMC including the reflex motion and center-of-mass (COM) displacement has been shown to be essential to reproduce the observed properties of MW stellar streams using orbital modeling \citep{erkal19a, vasiliev20, shipp21}. Furthermore, in a recent study \cite{vasiliev21} illustrated the importance of the wake, the COM displacement, and the reflex motion have on inducing radialization in the orbits of massive satellites.  

In this study, we present a theoretical framework in which to understand the consequences of: 1) our non-inertial reference frame, owing to the ``reflex motion''and the displacement of the central density peak of the MW's halo; 2) the kinematics of particles in the DM dynamical friction wake and collective response  (see G20); 3) the selection bias on measurements of the kinematics of objects already identified in a planar configuration. 
Using tailored N--body simulations of the MW--LMC encounter, we illustrate that, as a result of these listed effects, the orbital poles of the simulated MW halo DM particles cluster near that of the LMC and the VPOS.

In Section \ref{sec:sims}, we briefly discuss the N--body simulation of the MW--LMC system from G19. %\citet{garavito-camargo19a}. 
In Section \ref{sec:VPOSsims}, we illustrate the clustering of the orbital poles of MW halo particles at the present-day, $\approx$ 50 Myr after the pericentric approach of the LMC. We additionally introduce toy models to explain the physical origin of this clustering in Section \ref{subsec:reflex}. Furthermore, we explore how planar configurations are biased towards orbital poles clustering in Section~\ref{sec:bias}.   

In Section \ref{sec:discussion}, we discuss the time dependence of the simulated clustering of orbital poles and we discuss these results in the context of planes of satellites around other galaxies and in cosmological simulations. Finally, we conclude in Section \ref{sec:conclusions}.

\section{N--body Simulation Parameters}
\label{sec:sims}

\begin{table}
  \centering
  \begin{tabular}{l c c}
  \hline
  \hline
    \textbf{Component} & \textbf{Parameter} & \textbf{Value}  \\
    \hline
    DM halo & M$_{vir}$\footnote{In G19 there was an error and the values reported as $M_{vir}$ for the LMC are instead total halo masses $M_{halo}$.}, M$_{200}$, M$_{hern}$ $[10^{12}  \rm{M_{\odot}}]$ & $1.2, 1.03, 1.57$ \\
    & $R_{vir}, R_{200}$ [kpc] & $279, 208$ \\ 
    & $c_{vir}$, $c_{200}$ & 15, 11.2 \\
    & $a_{halo}$ [kpc] & 40.85 \\
    & DM halo particles & $10^{8}$ \\
    & Mass per particle [M$_\odot$] & $1.57 \times 10^{4}$ \\
    \hline 
    Disk & M$_{disk}\ [10^{10}$M$_{\odot}]$ &  5.78 \\
    & Disk scale length $r_a$ [kpc] &  3.5\\
    & Disk scale height $r_b$ [kpc] &  0.5\\
    & Disk particles & 1382310 \\
    \hline
    Bulge & M$_{bulge}\ [10^{10} \rm{M_{\odot}}]$ & 1.4 \\
    & scale length $a_{bulge}$ [kpc] & 0.7 \\
    & Bulge particles & 335220 \\
  \hline
   LMC  & $M_{halo}$, $M_{vir}$ $[10^{10}  \rm{M_{\odot}}]$ & 18, 11.1\\
   & $a_{LMC}$[kpc] & 20\\\
   & $R_{vir}$[kpc] & 148\\
  \hline
  \hline
  \end{tabular}
  \caption{\textbf{MW and LMC model parameters}. MW Models 1 and 2 have the same parameters, but are initialized with different anisotropy profiles. These are the same simulations as those presented in G19 but including updated values for the virial masses ($M_{vir}$) of the LMC. The simulations presented here correspond to simulations \#3 and \#7 in G19.}
  \label{tab:sims-models}
\end{table}

In this analysis, we use the fiducial N--body simulation of the MW--LMC encounter presented in G19. The main properties of this simulation are summarized in Table~\ref{tab:sims-models}. 
The simulation has a mass resolution of 
$m_p=1.5\times 10^{4} M_{\odot}$ and is run with Gadget-3 \citep{Springel08}. This simulation utilizes a MW halo with a radially biased anisotropy parameter, $\beta(r)=-0.15-0.2
\frac{dln \rho(r)}{dln r}$, and an LMC with a Hernquist halo mass \citep{Hernquist92} at infall of 1.8$\times 10^{11} M_{\odot}$. 
 Both halos of the MW and the LMC were initialized 
using GalIC \citep{Yurin14} with a Hernquist DM halo profile.
 The MW also includes a live bulge and disk. For more details we refer the reader to G19. 

\begin{figure}
    \centering
    \includegraphics[scale=0.6]{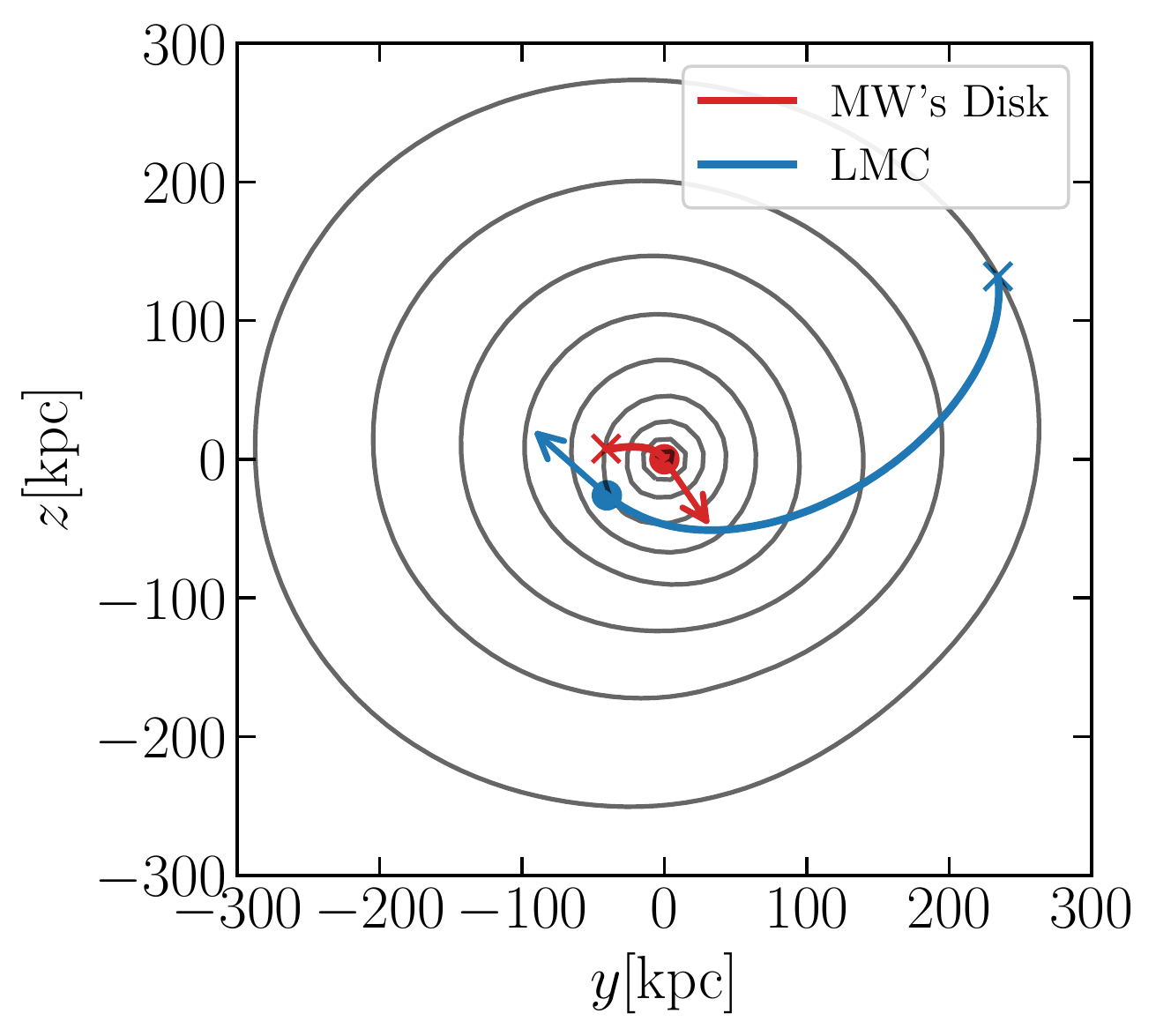}
    \caption{Isodensity contours for the simulated present-day DM halo of the MW in the Galactocentric $y$--$z$ plane, which is approximately the orbital plane of the LMC. The contours illustrate the distortions to the MW's halo owing to the LMC's orbit, but LMC particles are not included. After the LMC's infall, spherical symmetry is broken, causing the isodensity contours of the host DM halo to be distorted (see G20 for further details). The center of mass position of the MW's disk and the LMC's halo $\approx 2$ Gyr ago are marked by a red and blue \textit{x}, respectively, and their present-day locations are marked with circles. The present-day location of the MW disk (red circle) is no longer coincident with the centroid of the outer isodensity contours (the red x).
    The solid lines show the trajectories of the LMC and the MW disk, while the arrows show the corresponding present-day velocity vector. The inner halo is moving in the opposite direction as the LMC, while the outer halo is not. As a result to an observer in the disk the outer halo is co-orbiting with the LMC disk. The present-day velocity signature of this motion is referred to as ``reflex motion"}. 
    \label{fig:sketch}
\end{figure}

The simulation was run for $\approx$2 Gyr, and follows the evolution of 
the LMC on its first infall into the MW, accounting for the perturbations induced by the LMC to the DM distribution of the MW. Figure~\ref{fig:sketch} illustrates isodensity contours for the present-day DM distribution of the MW, after the arrival of the LMC. 
At the present time, the simulated LMC is just past orbital pericenter, and its current position and velocity vector is within 2$\sigma$ of that observed \citep{Kallivayalil13}. 

The MW disk also moves in response to the LMC \citep{Gomez15}. The disk motion is depicted by the red arrow in Figure~\ref{fig:sketch}. The disk moves up to $\approx$40~kpc over the past 2 Gyr (from the red x to the red circle).  As observers in the disk, we should see the reflex of this motion in observations of stars in the halo \citep{Petersen20b, Erkal20c}. In G19, this MW--LMC simulation was found to maximize the halo response, including the amplitude of the DM dynamical friction wake that trails the LMC.

We also study the properties of an isolated MW (no LMC) with these same halo properties for comparison. This isolated MW model is also run for 2 Gyr, allowing us to track particles over this same period in time, but without perturbations from the LMC.  We examine the impact of the assumed LMC infall mass  on our findings in Discussion Section~\ref{sec:tracers}. 

\section{The Clustering of Orbital Poles in Tailored N--body Simulations of the MW--LMC System}
\label{sec:VPOSsims}

\begin{figure*}[ht]
    \centering
    {\Large Density of orbital poles for DM halo particles}\par\medskip
    \includegraphics[scale=0.56]{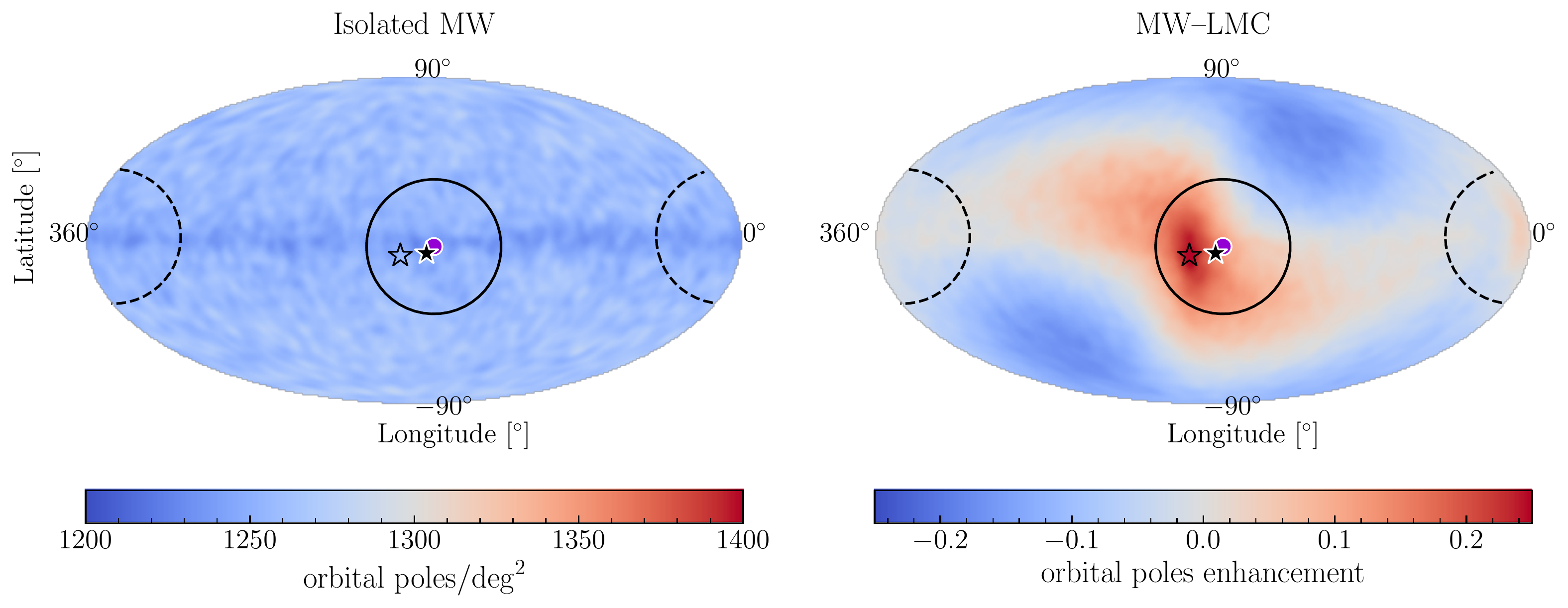}
    \caption{\textbf{Left panel:} All-sky Mollweide projection illustrating the distribution of orbital poles (i.e. the angular momentum vector of each particle projected on the sky) for MW DM halo particles at Galactocentric distances 50 $< r_{GC} < $ 400 kpc in the isolated MW simulation (no LMC, left panel) at the present time. 
    The orbital poles are computed with respect to the Galactic center.  The density of poles is uniform across the sky$^{2}$. \textbf{Right panel:} The ``orbital poles enhancement," i.e. the density contrast of orbital poles in the MW--LMC simulation relative to the isolated MW simulation, as defined in Equation~\ref{eq:enhancement}. The density of poles is no longer uniform and there is a $\approx20\%$ peak enhancement with respect to the isolated MW simulation along a sinusoidal-shape, centered on the simulated LMC's orbital pole (open black star). The peak enhancement ($\approx25\%$) is centered on the orbital pole of the simulated LMC, and near the mean normal to the VPOS purple dot ($(l,b) = (169.3^{\circ}, -2.8^{\circ})$). The true, observed LMC's orbital pole (filled black star) deviates somewhat from the simulated LMC, but is consistent within the observational errors. The black solid circle encompasses 10\% of the sky around the assumed VPOS (an opening angle of $36.87^{\circ}$). Note that no LMC particles were included in these figures, as we aim to understand changes in the kinematics of objects in orbit about the MW prior to the LMC's infall.
    }
    \label{fig:MWLMCsims}
\end{figure*}

As new substructures have been discovered and as improved observational measurements have become available, consistent orbital planes have been fit to the observed kinematics of satellites, streams, and globular clusters around the MW  \citep[e.g.,][]{Kroupa2005, Metz2008,Metz2009,Pawlowski2012, Pawlowski2013,Pawlowski2015, Santos20, Pawlowski2020, Li21}. Here, we adopt the definition of the normal to the VPOS as in \citet[][see  Figure~\ref{fig:MWLMCsims}]{Pawlowski2013}.
Co-rotating satellites then have orbital poles\footnote{Computed as $\hat{\bm{\jmath}} = \frac{\vec{r} \times \vec{v}}{|\vec{r} \times \vec{v}|}$, using the 3D position ($\vec{r}$) and velocity ($\vec{v}$) vectors in a frame centered on the MW today. The resulting direction of the angular momentum vector, $\hat{j}$ is then transformed into Galactocentric spherical coordinates $(longitude, latitude)$ in degrees.} similar to the normal to the VPOS and counter-rotating satellites have orbital poles that are offset by $180^{\circ}$ in Galactic longitude (also referred to as the anti-VPOS). The orbital pole of the simulated and real LMC is within the region encompassing 10\% of the sky around the assumed orbital pole of the VPOS.

In the following, we examine the distribution of the orbital poles in our isolated MW and MW--LMC simulations in relation to the observed orbital poles of the LMC, VPOS and anti-VPOS.

%%%%%%%%%%%%%%%%

\subsection{Orbital Poles of Particles in the MW--LMC Simulation vs. the Isolated MW Simulation}
\label{sec:op_MWLMC_isoMW}

To determine the role of the LMC in the observed clustering of satellite orbital poles, we first examine the distribution of orbital poles computed for all MW DM particles (50-400 kpc) in the simulation output corresponding to the present-day. The left panel of Figure~\ref{fig:MWLMCsims} illustrates the density of orbital poles for MW halo particles in the isolated MW simulation. This plot represents the distribution of Galactocentric
azimuthal (longitudinal)
and polar (latitudinal) angles of the projected location of the angular momentum vector of each DM particle on the sky. In the isolated case, the density of orbital poles is approximately homogeneous across the sky  \footnote{There is a small underdensity in orbital poles at Lat=0$^{\circ}$ caused by the initial conditions generated with GalIC.}, as expected for a pressure supported system in equilibrium. This is the initial state of the MW halo before the LMC arrives.

The right panel illustrates the enhancement of orbital poles in the MW--LMC simulation with respect to the isolated MW simulation. The enhancement is defined as: 

\begin{equation}
   \delta \rho_{iso} = \dfrac{\rho_{\rm{poles}}(MW+LMC)}{\rho_{\rm{poles}}(MW)}-1.
\label{eq:enhancement}
\end{equation}
 
We also defined the enhancement between orbital poles overdensities and underdensities as:  
 
\begin{equation}
   \delta \rho_{max} = \delta \rho_{iso, max} - \delta \rho_{iso, min}
\label{eq:relative_enhancement}
\end{equation}
 
Where $\delta \rho_{iso, max}$ and $\delta \rho_{iso, max}$ are the maximum and minimum values of $\delta \rho_{iso}$. As such $\delta \rho_{max}$, is the maximum enhancement of orbital poles.

For the MW--LMC simulation (right panel), the orbital poles
cluster near the orbital pole of the simulated LMC and the normal 
direction of the VPOS. Note that the simulated orbital pole 
differs by $\approx 5^{\circ}$ from the  observed orbital pole of the LMC, but is consistent within observational errors. The clustering results in an increased density of orbital poles near the LMC. The peak 
enhancement is $\delta \rho_{iso} \approx 25\%$, while the maximal enhancement is $\delta \rho_{max} \approx 40\%$. Furthermore an 
enhancement of $\delta \rho_{iso}\approx10\%$ on average along a sinusoidal pattern is 
present in the projected distribution of orbital poles today. 
There are correspondingly two under-dense regions in the Northeast and the Southwest, each $\delta \rho_{iso} \approx 15\%$ less dense as compared to the density of orbital poles in the isolated MW. The sinusoid also encompasses the anti-VPOS region (dashed black circle).

\subsection{The Clustering of Satellite Orbital Poles with Galactocentric Distance} 
\label{sec:radial}

\begin{figure*}
    \centering
    \includegraphics[scale=0.7]{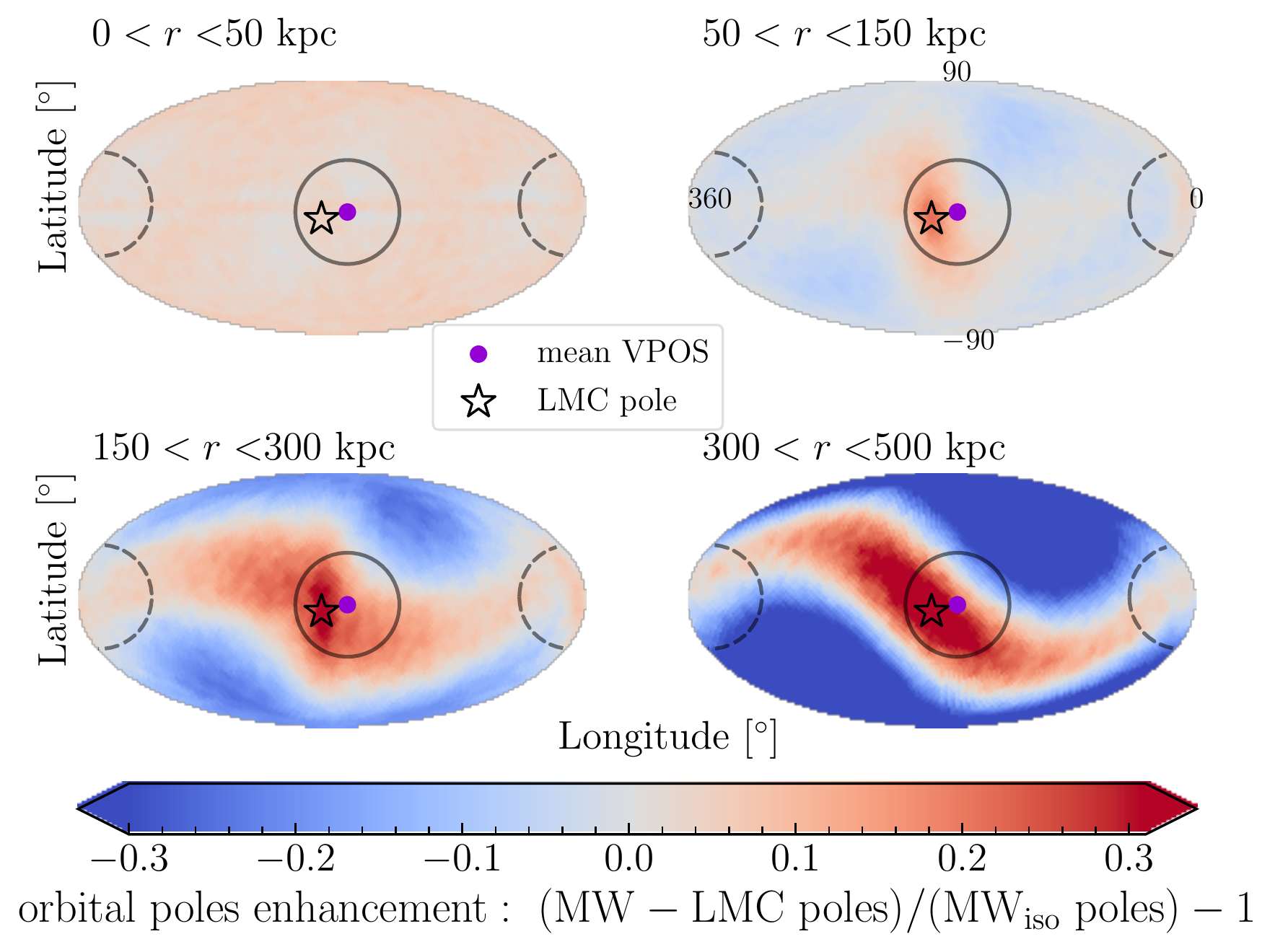}
    \caption{Enhancement in the density of orbital poles as a function of Galactocentric distance in the MW--LMC simulation. The enhancement is computed as in Equation~\ref{eq:enhancement}. In the innermost regions ($<$ 50 kpc), the density of orbital poles is approximately uniform. Moving outwards to the 50-150 kpc panel (top right), a
density peak near the projected orbital pole of the VPOS (purple point) and simulated LMC's orbital pole (open black star) is evident. Also, the underdense regions in the Northeast and Southwest, seen in Figure~\ref{fig:MWLMCsims}, start to appear. In the outermost
distance bins (150-300 kpc and 300-500 kpc), the distribution of poles continue to fill in the sinusoidal shape seen in Figure~\ref{fig:MWLMCsims}, stretching across Galactic Longitudes of 0$^{\circ}$-360$^{\circ}$. The enhancement of poles strengthens as a function of Galactocentric distance with maximal values of $ \delta \rho_{iso}\approx15\%, 25\% 40\%$ (note that the color is saturated and extents beyond the range shown in the colorbar), at distances of $50-150, 150-300$, and $300-500$ kpc, respectively. The values are higher when comparing to the background (blue regions) in the MW--LMC simulation, reaching up to $ \delta \rho_{max} \approx  30\%, 50\%, 80\%$ in the same distance ranges.
    \label{fig:MWLMC_distance}}
\end{figure*}

In Figure~\ref{fig:MWLMC_distance} we repeat the previous exercise and plot the enhancement of orbital poles for particles within the specified range of Galactocentric distances. 
Within 50 kpc, there is no coherent signature amongst the orbital poles of halo particles. This is consistent with expectations from G20 where the center of mass of the inner halo ($<$ 30 kpc) is roughly coincident with the center of mass of the disk. As such, changes to the reference frame of the observer would be minimal. This is also consistent with recent findings that many globular clusters and stellar streams may not be aligned with the orbital pole of the LMC \citep{Riley20}, as they are mostly within 30 kpc. {\color{black}This has also been shown for BHB stars within 50 kpc \citep{Deason11a}}. 

In contrast, the orbital poles of DM particles become clustered around that of the LMC and VPOS at radii larger than 50 kpc, where the majority of satellites reside. At Galactocentric distances of 50-150 kpc, the highest density is concentrated around the normal to the VPOS/the simulated orbital pole of the LMC. At radii larger than 150 kpc, the sinusoidal pattern seen in Figure~\ref{fig:MWLMCsims} forms.
Together, these two patterns indicate an \emph{apparent} observed preferred direction in the angular momentum vectors 
of DM particles that strengthen as a function of distance beyond 50 kpc.We expect this signal to be exhibited by all objects in orbit about the MW. In the following we seek to explain the physical origin of this effect. 

%%%%%%%%%%%%%

\section{Understanding the Simulated Distribution of Orbital Poles}
\label{subsec:reflex}

In the following sections, we illustrate that the clustering of orbital poles seen in our fiducial MW--LMC simulation (Figures~\ref{fig:MWLMCsims},\ref{fig:MWLMC_distance}) 
is primarily the result of three effects: 1) the non-inertial reference frame of the MW's disk as a result of the shift in phase space of the inner halo (reflex motion and displacement of the inner halo); 2) the kinematics of particles located in the DM dynamical friction wake and collective response  induced by the LMC; and 3) the bias in the distribution of orbital poles caused by the selection of tracers in non-spherical distributions.

\subsection{The impact of a non-inertial reference frame}

A non-inertial reference frame for observers in the disk of the MW relative to the halo stems from the fact that the barycenter of the MW's halo has been counter-rotating with respect to the LMC's orbital plane for the past $\approx 2$ Gyrs. As such, it has an impact on the present-day kinematics of the halo in the following ways: 

\begin{enumerate}
    \item The relative displacement of the central density peak of the MW's disk and inner halo
    with respect to the center of mass (COM) of the MW's outer halo ($>$ 30 kpc) varies as a function of Galactocentric distance \citep[][G20]{Gomez16}. This causes a large overdensity in the DM/stellar halo in the northern hemisphere, which we refer to as the collective response  \citep[see G19,G20,][]{conroy21}.
    \item The velocity of the MW disk with respect to the velocity of the outer halo ($>$ 30 kpc) also varies as a function of Galactocentric radius, where we observe the ``reflex" of this motion \citep[e.g.,][]{Gomez15, Erkal18b, garavito-camargo19a, Peterson20,Erkal20c, Cunningham20}.
  
\end{enumerate}

These two effects are illustrated in Figure~\ref{fig:sketch}. Together, these effects shift our reference-frame in both position and velocity relative to the outer halo, where the amplitude of these shifts varies as a function of Galactocentric radius. As a consequence of these two effects, the reference frame of an observer in the disk cannot be approximated as inertial with respect to objects in orbit throughout the halo.

\subsubsection{Center of Mass Displacement}
\label{sec:com_displacement}

\begin{figure}[h]
    \centering
    \includegraphics[scale=0.6]{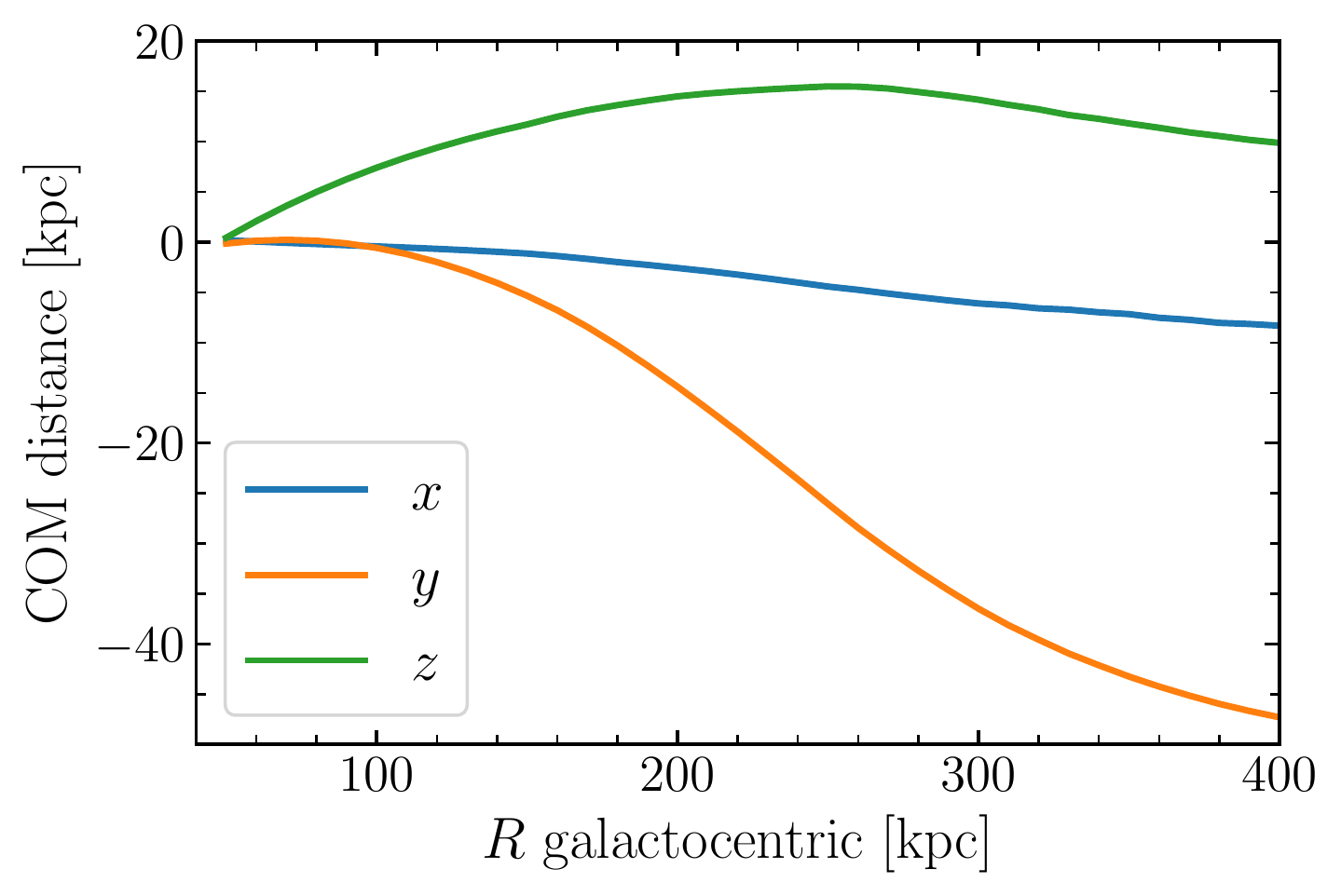}
    \includegraphics[scale=0.6]{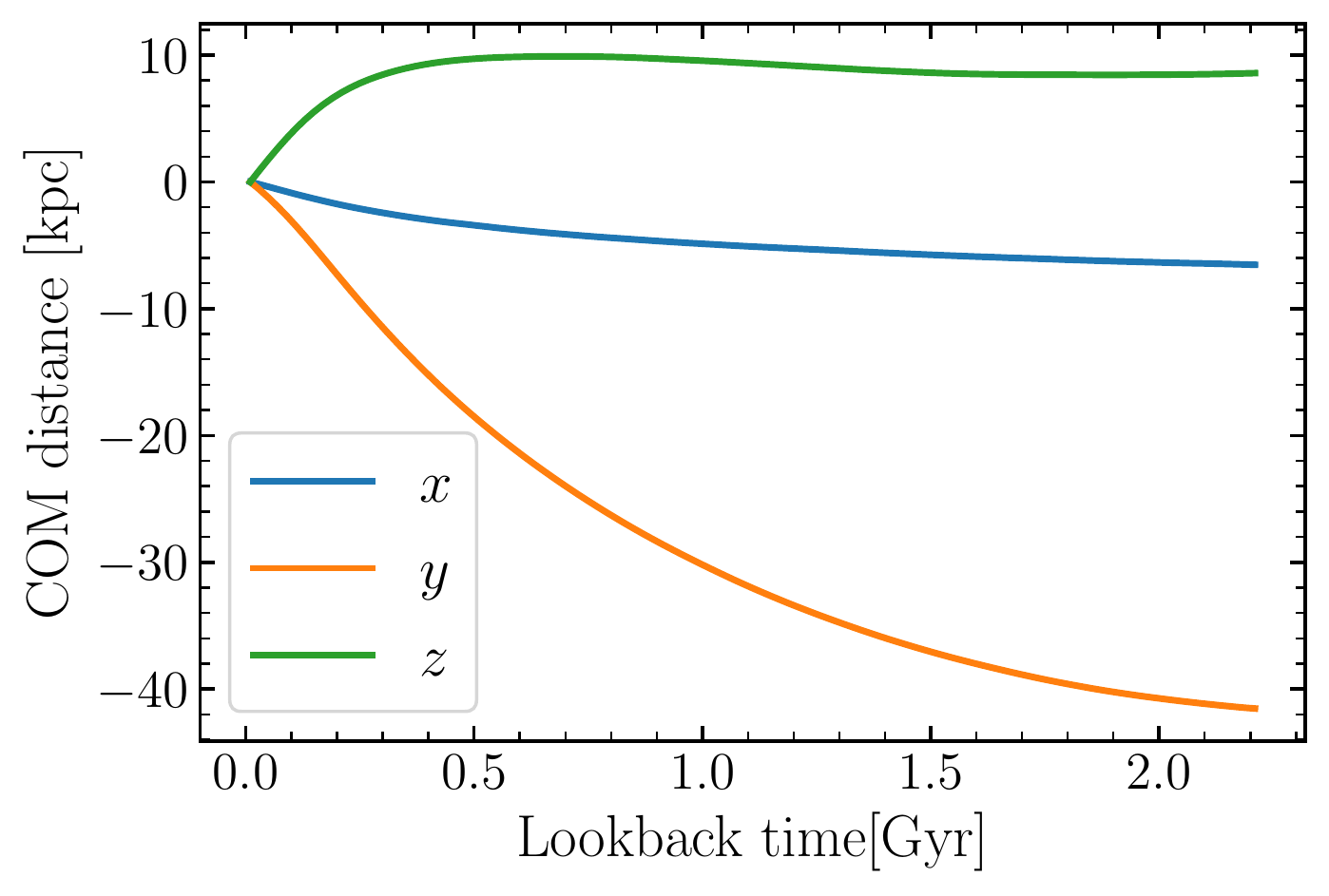}
    \caption{\textbf{Top panel:} Position vector of the COM of the DM halo of the MW in the MW--LMC simulation as a function of Galactocentric distance. The halo COM is computed in Galactocentric spherical shells of 10 kpc thickness using only MW halo particles.  \textbf{Bottom panel:} The change in position of the COM of the MW's disk throughout the course the MW--LMC simulation, where the present time is at 0 Gyr. The shift in disk's COM is computed relative to its initial COM location prior to the infall of the LMC (i.e. 2.5 Gyr ago). The change in the disk position is correlated with the change in the COM of the halo (top panel).} 
    \label{fig:com_displacement}
\end{figure}

We first explore the impact of the COM displacement induced by the LMC on the density of orbital poles of particles in the halo. As described in \citet{Gomez15}, the COM of the MW disk and halo within 30 kpc is displaced by 20-50 kpc (depending on the LMC's mass) over the last 2 Gyrs due to the passage of the LMC (see the bottom panel of Figure~\ref{fig:com_displacement}). This shift in position is computed with respect to the location of the MW's disk prior to the infall of the LMC. 

However, because the halo is not a solid body, the displacement of the halo is larger in the inner regions than in the outer regions of the halo. Consequently, the COM of the halo changes as a function of distance. 
Here we quantify the effect caused by such spatial offsets throughout the halo in the density of orbital poles.

We compute the COM of the MW's halo in Galactocentric spherical shells of 10 kpc thickness (top panel of Figure~\ref{fig:com_displacement}). The COM position vector of each shell is plotted with respect to the location of the present-day COM of the MW disk. The COM of the halo begins to deviate from that of the disk beyond 30 kpc, in agreement with previous studies. Note that both the $\hat{y}$ and $\hat{z}$ component of the COM position vector are most significantly affected. This is expected since the orbit of the LMC roughly lies in the $y-z$ plane. 

The bottom panel of Figure~\ref{fig:com_displacement} illustrates the motion of the MW's disk in the last 2.5 Gyrs with respect to its initial location, prior to the infall of the LMC. As in the top panel, the disk of the MW moves primarily in the $y-z$ plane. There is a clear correspondence between the COM displacement and the disk displacement in the last 2.5 Gyrs (top and bottom panel in Figure~\ref{fig:com_displacement}). The location of the disk COM at the start of the simulation (2.5 Gyr ago) is similar to the COM of the MW's DM halo at distances greater than 200 kpc, indicating that the outer halo lags the displacement of the inner halo.

Note that within 100 kpc, the COM does not change significantly. 
The DM dynamical friction wake induced by the LMC traces the LMC's orbit, which is located mainly at positive values of $y$  at distances of 50-150 kpc (see blue line in Figure~\ref{fig:sketch}). Therefore, overdensities in the halo owing to the displacement of the COM (collective response) are counterbalanced by the overdensity of the dynamical friction wake, which is located in the opposite region of the halo.
We will discus the impact of the dynamical friction wake in more detail in Section~\ref{sec:wake}. 

More generally, because we have not accounted for the LMC's particles, what we have computed here is not the orbital barycenter of the MW--LMC system, but rather the centroid of only the MW's halo mass distribution as a function of distance. Deviations from spherical symmetry in the halo due to the COM displacement imprints a particular signal in the density of orbital poles. To first order, this can be quantified by performing a simple position shift in a spherical halo. Note that this is only an approximation, as shown in G20, the distortions to the halo are highly asymmetric. This effect is not captured in this analysis.

\begin{figure}[h]
    \centering
    \includegraphics[scale=0.7]{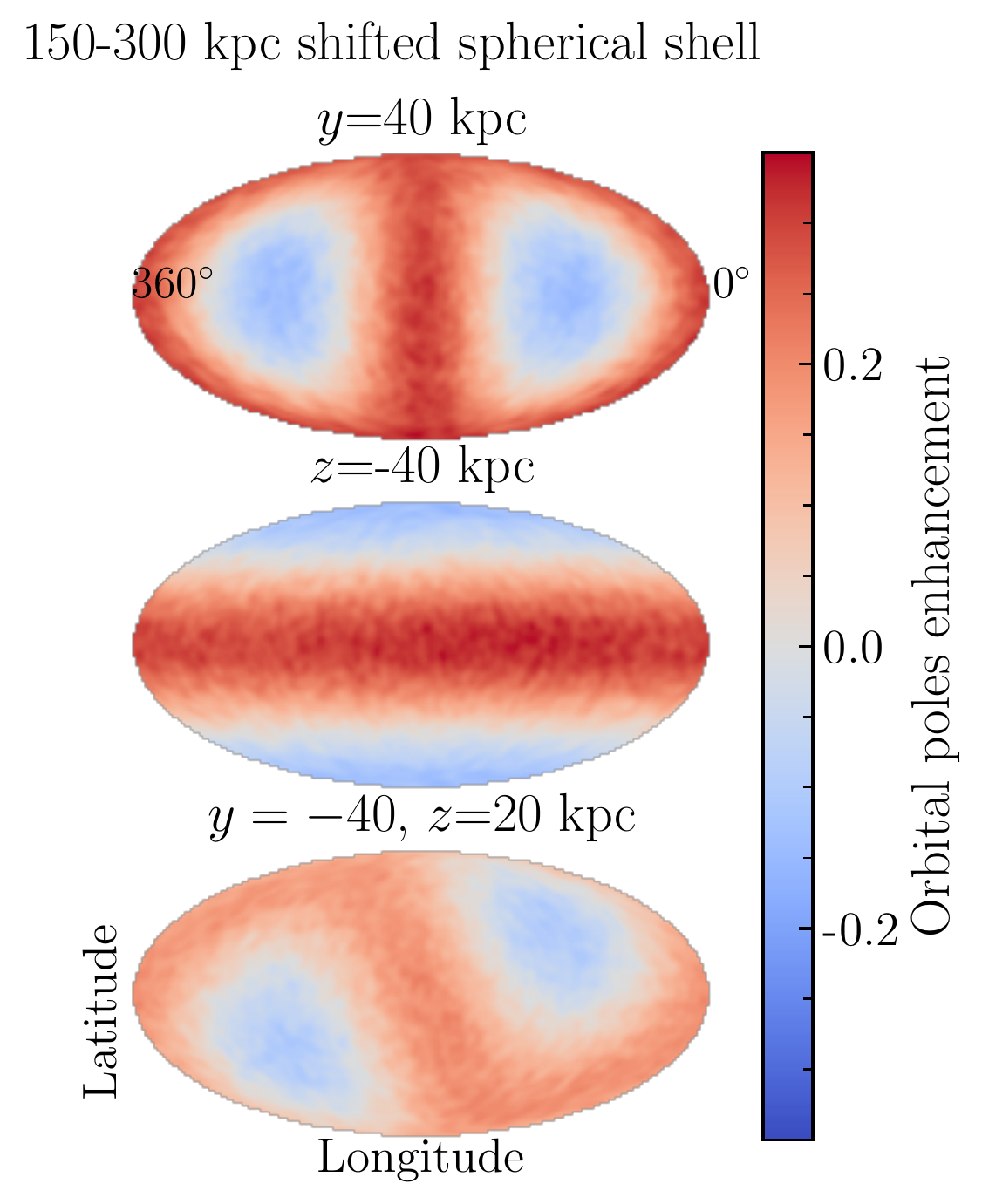}
    \caption{Enhancement of orbital poles projected on the sky, computed using DM halo particles in the isolated MW halo simulation whose centroid is offset from the observer in three different directions, as listed.  We do not change the kinematics of the particles. Particles are selected within a spherical shell between 150-300 kpc, centered on the observer. The resulting spherical shell has over/under densities owing to the offset. The resulting orbital poles are distributed along a great circle on the sky whose inclination depends on the angle of the offset. The appearance of these patterns is entirely due to density inhomogeneities in the spherical shell as viewed by an offset observer.}
    \label{fig:com_shift_ex}
\end{figure}

In Figure~\ref{fig:com_shift_ex} we illustrate the enhancement in the density of orbital poles of DM halo particles in the isolated MW halo simulation (no LMC; 150-300 kpc), where the centroid of the MW halo is offset from the observer by the indicated vector. In other words, the position vector of each particle in the halo is shifted by the listed amount/direction, while the observer is located at the original centroid of the unperturbed halo. 

The top panel of Figure~\ref{fig:com_shift_ex}
shows the resulting enhancement of orbital poles when the halo is offset by +40$\approx$kpc along the $y-$axis from the observer. The resulting orbital poles are clustered in a great circle at Lon$=0^{\circ}, 180^{\circ}$. Similarly, if the halo is offset in the negative $z$-direction, the enhancement of poles seen by the observer is in a great circle at Lat$=0^{\circ}$ (middle panel). As such, any spatial off-set induces an enhancement of poles in a great-circle, where the orientation depends on the direction of the offset. The amplitude of the enhancement depends on the amplitude of the spatial off-set. In the case of the shift that most closely mimics displacement of the MW disk owing to the LMC (bottom panel, $y=-40$ kpc and $z=-20$ kpc), the resulting distribution of orbital poles is a sinusoidal pattern, as shown in Figure~\ref{fig:com_shift_ex}. However, a COM displacement alone cannot reproduce the clustering seen in the orbital pole enhancement of the MW--LMC simulation (Figure \ref{fig:MWLMCsims}).

\subsubsection{Reflex Motion}
\label{sec:reflex_motion}

\begin{figure*}
    \centering
    \includegraphics[scale=0.7]{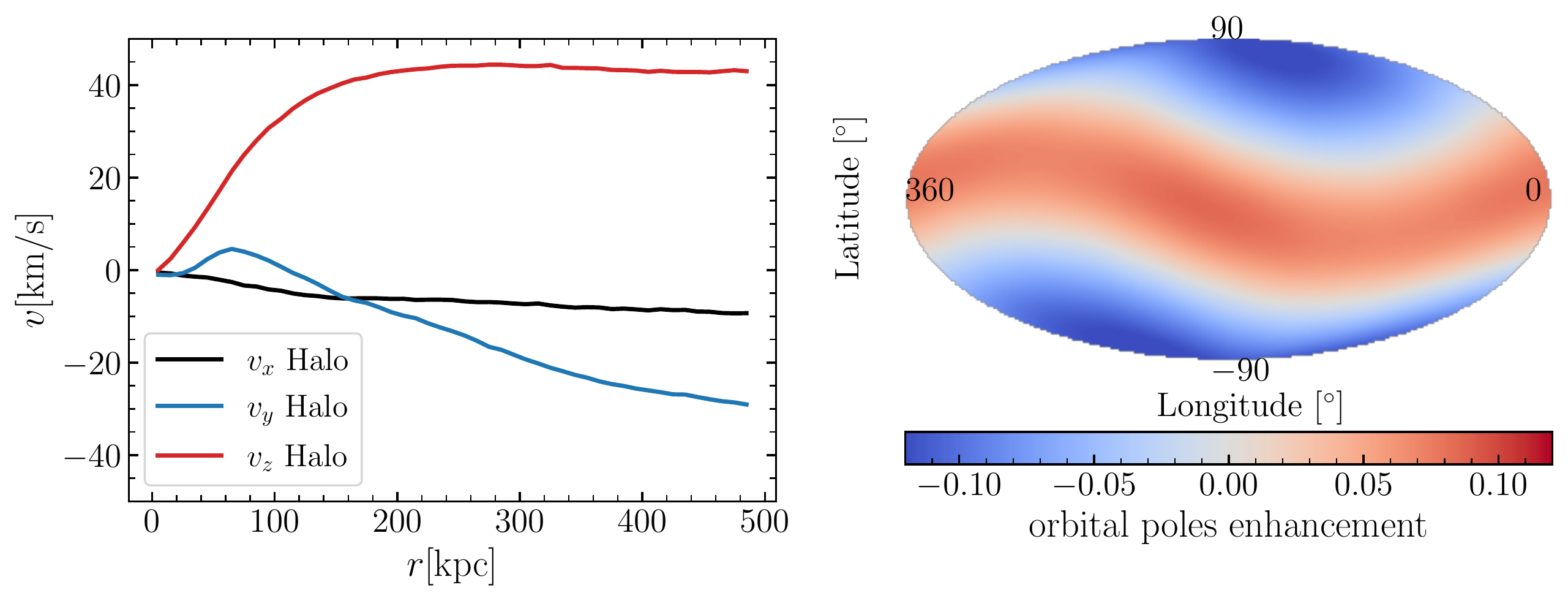}
    \caption{The impact of the present-day reflex motion on the orbital poles of DM halo particles. \textbf{Left panel:} The mean velocity vector of MW halo particles in the MW--LMC simulation computed within spherical shells and plotted per vector component as a function of distance. Velocities are computed with respect to an observer at the Galactic center. This is the ``reflex motion" an observer in the MW's disk would see as a function of distance. \textbf{Right panel:} The enhancement of orbital poles of particles in the isolated MW halo simulation, but where velocity shifts were introduced to all particles according to the left panel (no displacement is introduced). 
    The sinusoidal-shape seen in the MW--LMC simulation (Figure~\ref{fig:MWLMCsims}) is reproduced in the isolated MW model through this process. However, the sinusoid does not exhibit stronger clustering near the orbital poles of the VPOS or the LMC. Reflex motion alone cannot reproduce the signal seen in the MW--LMC simulation.}
    \label{fig:sin-shape}
\end{figure*}

\begin{figure*}[h]
    \centering
    \includegraphics[scale=0.6]{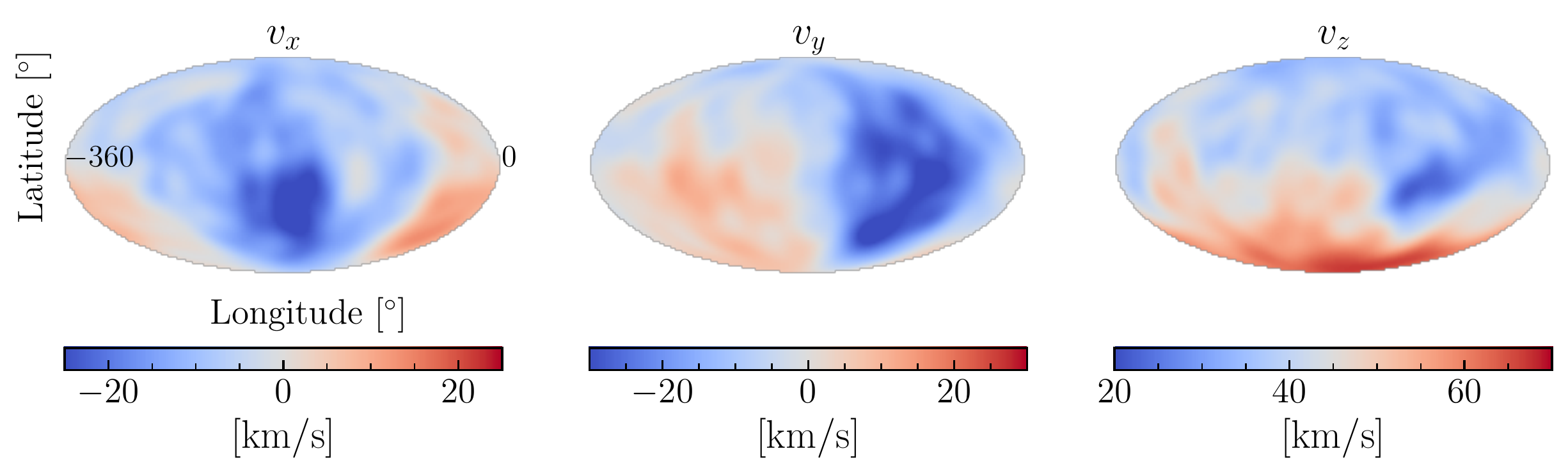}
    \caption{Angular variations in the velocities of the DM particles in the MW--LMC simulation in a spherical shell from 150-160 kpc, projected on the sky. Each panel shows a component of the velocity vector, where the particle velocities are averaged over 2 square degrees bins. The resulting variations in velocities were caused by the passage of the LMC as described in \citep{garavito-camargo19a, Peterson20, Cunningham20} and later discovered by \citep{Petersen20b, Erkal20c}. Most of these features are associated with the ``reflex motion", but there are also signatures from the dynamical friction wake in the south (see G19).
    }
    \label{fig:angular_reflex}
\end{figure*}

Having explored the effect of the COM displacement of the halo, we now focus on the effect of the reflex motion we observe as a consequence of that displacement. Like the COM displacement, the relative motion of the disk with respect to the halo varies as a function of Galactocentric distance, 
polar, and azimuthal angle. \citep{garavito-camargo19a, Cunningham20, Erkal20}.   
Using our fiducial MW--LMC simulation, we compute 
the mean velocities of the MW halo particles within spherical shells with respect to the COM of the disk of the MW at the present-day. We plot the resulting velocity vector as a function of distance, computed with respect to an observer at the Galactic center in the left panel of Figure~\ref{fig:sin-shape}; this is the ``reflex motion''.

The amplitude of the mean reflex velocity increases as a function of distance and is largest in the $v_z$ component. As discussed in previous work \citep{garavito-camargo19a,Erkal20, Peterson20, Cunningham20, garavito-camargo2020}, the reflex motion is most pronounced at distances beyond 30 kpc, where the halo's dynamical times are larger than the orbital period of the LMC.

To study the impact of this motion on the orbital poles of particles, we add this radially varying, reflex velocity vector to each halo particle in the isolated MW simulation (no LMC), according to their Galactocentric radius. The resulting enhancement in the distribution of orbital poles of DM particles in the MW halo also exhibits a sinusoidal pattern (right panel of Figure~\ref{fig:sin-shape}). This pattern has no preferred central clustering but does reproduce the overall pattern seen in the right panel of Figure~\ref{fig:MWLMCsims}. By introducing a preferred direction of motion (reflex velocity vector) we preferentially enhance orbital poles along a great circle projected on the sky, just as we showed in the previous section for the COM displacement. 

Thus far, we have computed the mean reflex velocity vector as an average across an entire spherical shell. However, because the LMC orbit has a preferred direction, the reflex velocity vectors of particles vary not only in radius, but also in angle on the sky. There are, in particular, bulk motions induced by the dynamical friction wake that cannot be accounted for properly in an all-sky average \citep[GC19,][]{Cunningham20}. In Figure~\ref{fig:angular_reflex}, we plot the all-sky projection of each component of the velocity vector of particles within 150-160 kpc, averaged within angular bins of 2 square degrees. We have binned the halo out to 300 kpc, but for illustrative purposes we show the velocity map at 150-160 kpc. The angular variations in reflex velocity can vary by more than 40 km/s, making it essential to utilize such maps to correct for the reflex motion, rather than using all-sky averages. 
In the next section we will utilize these angular corrections in addition to the COM displacement to study the impact of our non-inertial reference frame on the enhancement of orbital poles seen in the MW--LMC simulation.

\subsubsection{Reflex Motion \& Displacement: A non-inertial reference frame}

\begin{figure*}[h]
    \centering
    \includegraphics[scale=0.65]{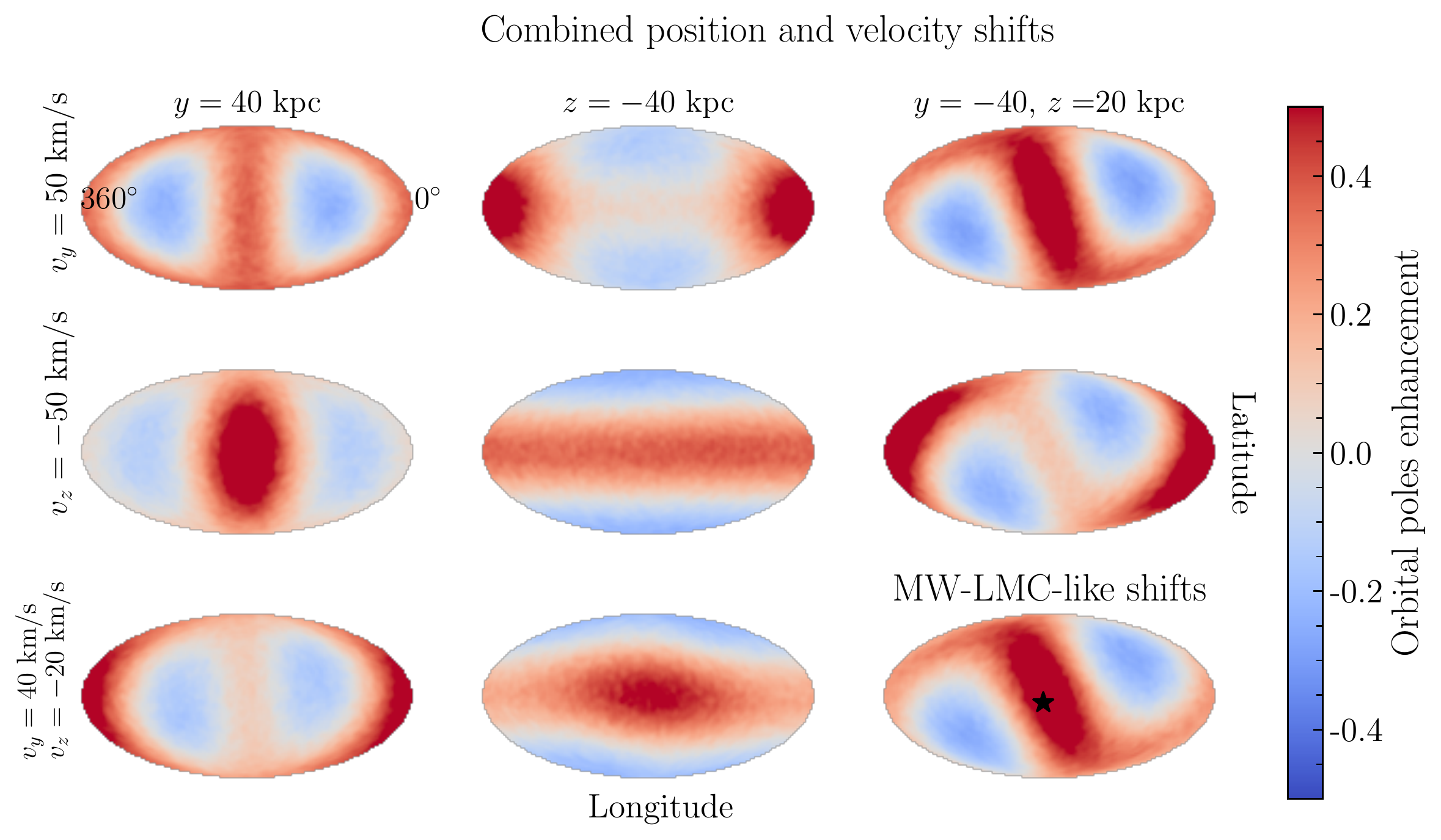}
    \caption{Enhancement in the all-sky density of orbital poles after adjusting the position and velocity vectors of DM halo particles in an isolated MW halo simulation (no LMC) using the listed COM displacement (columns) and reflex motion (rows). The shifts in particle positions and velocities are made to all particles in the entire isolated MW halo. To construct the images, particles are selected within 150-300 kpc of the original Galactic center of the unperturbed MW. The bottom right panel most closely mimics the expected COM displacement and reflex motion of the MW halo owing to the LMC at these distances. 
    The combined test position and velocity adjustments result in enhancements of the distribution of orbital poles following great circles on the sky. Interestingly the inclusion of both effects also cause non-homogeneous clustering, which is not seen when only positions or velocities are adjusted (Figures~\ref{fig:com_shift_ex} or \ref{fig:sin-shape}). The exception is in cases where the adjustments in positions and velocities are made in the same direction (top left and middle panels).}
    \label{fig:poles_combined_shif}
\end{figure*}

In Sections~\ref{sec:com_displacement} and \ref{sec:reflex_motion} we demonstrated that shifts in either the position or velocity of halo DM particles can induce an enhancement of orbital poles along a great circle on the sky. We now explore the combined effect of a shift in both positions and velocities. 

\begin{figure*}[h]
    \centering
    \includegraphics[scale=0.6]{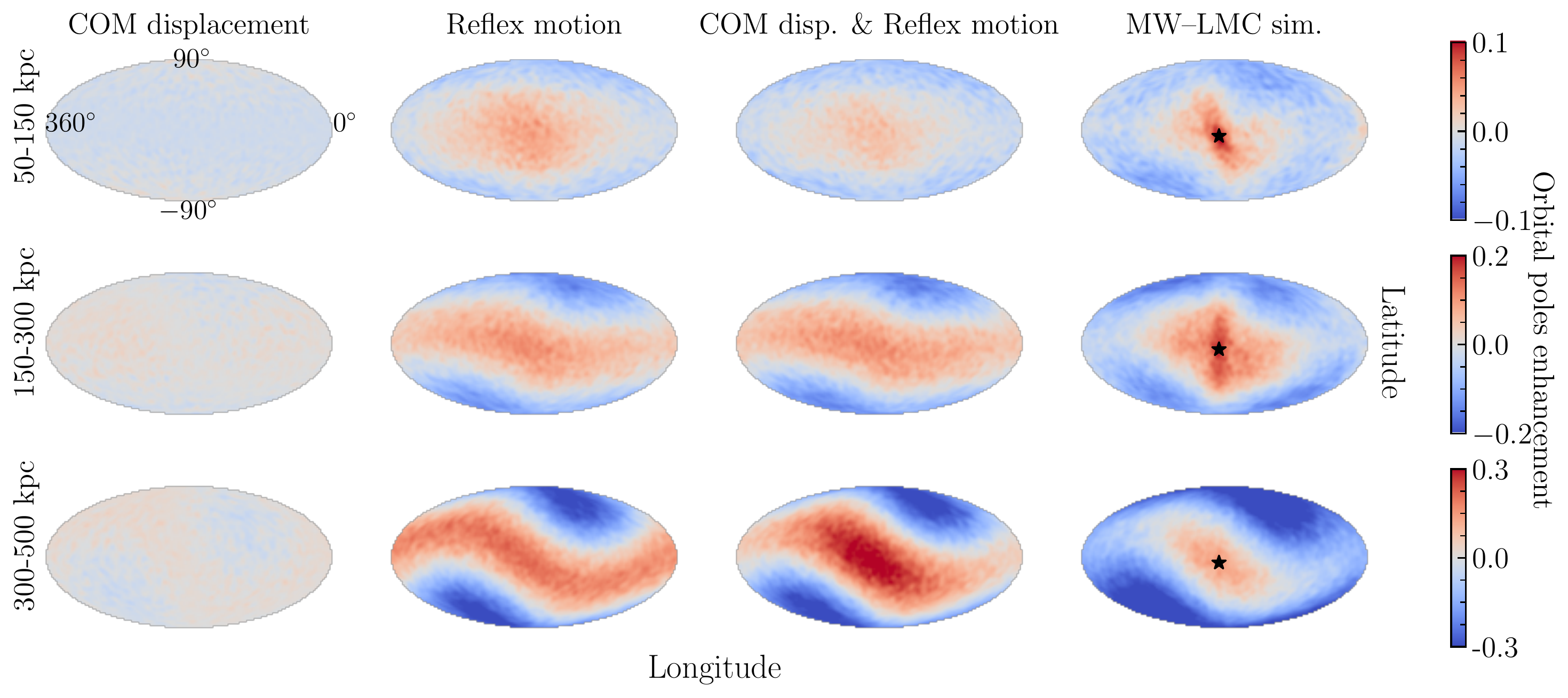}
    \caption{Enhancement in the density of orbital poles resulting from the radially varying COM displacement (first column) and reflex motion (second column) caused by the LMC.
    For particles in each shell, positions are adjusted according to the COM displacement at the distance of the shell, as in Fig~\ref{fig:com_displacement} (top panel).
    The same strategy as Figure~\ref{fig:angular_reflex} is employed to correct particles in shells for the reflex motion at that distance, accounting also for the angular location of the particles on the sky (second column from left). The third column indicates the combined effect of the COM displacement and reflex motion. The fourth column indicates the enhancement in the density of orbital poles of MW halo particles in the actual MW--LMC simulation. The black star shows the orbital pole of the simulated LMC. Radially varying COM displacements and radially + angular variations in reflex motion can reproduce the overdensity of orbital poles along a great circle, but cannot reproduce the high density peak in the MW--LMC simulations that is co-incident with the orbital pole of the LMC. }
    \label{fig:poles_summary}
\end{figure*}

To build intuition to understand these dual effects in various directions and components, we start with the isolated MW halo simulation and adjust the velocity of all halo particles between 150-300 kpc using the following three vectors: $[0,50,0]$ km/s, 
$[0,0,-50]$ km/s, and $[0,40,-20]$ km/s. We then adjust the position of the halo particles in each of the three cases following the example in Figure~\ref{fig:com_shift_ex}, resulting in a total of 9 test halos. In all cases the observer is at the original centroid of the unperturbed halo. The halo with a combination of shifts of $[0, -40, 20]$ kpc and $[0,40,-20]$ km/s in positions and velocities is motivated by the motion of the MW's disk with respect to the outer halo at 300 kpc, as shown in Figures~\ref{fig:com_shift_ex} and \ref{fig:sin-shape}. 

Figure~\ref{fig:poles_combined_shif} shows the resulting enhancements in the density distribution of orbital poles for the selected DM particles in each of the 9 different halos, with respect to the original isolated halo. These tests do not yet account for the angular changes in the reflex motion on the sky or the COM position behavior with distance.

We see that, overall, the enhancement of orbital poles follows a great circle. However, the enhancement is no longer homogeneous. There is a gradient in the overdensity along the great circle, with a clear maximum and a minimum separated by $180^{\circ}$. However, if both the velocity and position shifts were made in the same direction, the enhancement is still homogeneous along the great circle (top right and middle panels). The effect is maximum if the shifts are performed in perpendicular directions (middle left, upper middle, and bottom right panels), which is similar to the real case of the MW--LMC system (bottom right panel). 

These results illustrate that a shift in positions and velocities can induce co-rotating and counter-rotating planar structures. In particular, in the bottom right panel that resembles the MW--LMC system, the shifts induce a large enhancement of orbital poles near the simulated LMC's pole (black star).

We now turn our attention towards more accurately mimicking the COM position displacement and reflex motion exhibited in the MW--LMC simulation. Here we use the radially varying COM position vectors computed in Section~\ref{sec:com_displacement} (top panel of Figure~\ref{fig:com_displacement}) and the radially varying, angular reflex motion vectors computed in Section~\ref{sec:reflex_motion}. We perform the shifts using DM particles in the simulation, following the same procedure as in the test cases.

The results of only COM position displacement or reflex motions corrections are applied is in the first and second column of Figure~\ref{fig:poles_summary}, respectively. The resulting maps of the enhancement of orbital pole densities if both corrections are applied are shown in the third column from the left of Figure~\ref{fig:poles_summary}. Each row illustrates the results for particles selected within the marked distance range. For comparison, we show the actual orbital poles of DM halo particles measured in the MW--LMC simulation in the fourth column. 

Within 50-150 kpc (top row), in the MW--LMC simulation, the orbital poles of DM particles cluster around the orbital pole of the LMC and VPOS. The reflex motion alone can mimic some of this clustering behavior, however, the amplitude is not as strong and centralized as observed in the MW--LMC simulation. 

Between 150-300 kpc (middle row), the reflex motion and COM displacement, respectively, result in a cluster of orbital poles that is homogeneous along a great circle. 
This test indicates that the COM displacement alone is not strong enough to explain the signal in the MW--LMC simulations within 300 kpc. In part, due to the fact that we are implementing a first order approximation for the position shift (see Section ~\ref{sec:com_displacement}). 

Beyond 300 kpc (bottom row), the inclusion of both position and velocity shifts result in orbital pole distributions with the same shape as that observed in the MW--LMC simulation. 
However, the amplitude is larger, implying that the directions of the shifts in the velocity and position vectors are correct, but their amplitudes are not.

These results highlight the difficulty of quantifying the exact COM displacement. As discussed in Section~\ref{sec:com_displacement} the COM calculation can only be computed in spherical shells and it does not accurately take into account the spatial location and morphology of the halo deformations and the dynamical friction wake (see G20). 

In the MW--LMC simulations (right column) within 300 kpc, there is an additional, narrow overdensity that is coincident with the orbital pole of the LMC but does not seem to be explained by the position and velocity shifts. In the next section, we explore the nature of this feature, which we associate with the presence of the DM dynamical friction wake.

\subsection{Effect of the Dynamical Friction Wake and Collective Response }\label{sec:wake}

\begin{figure*}[h]
    \centering
    \includegraphics[scale=0.41]{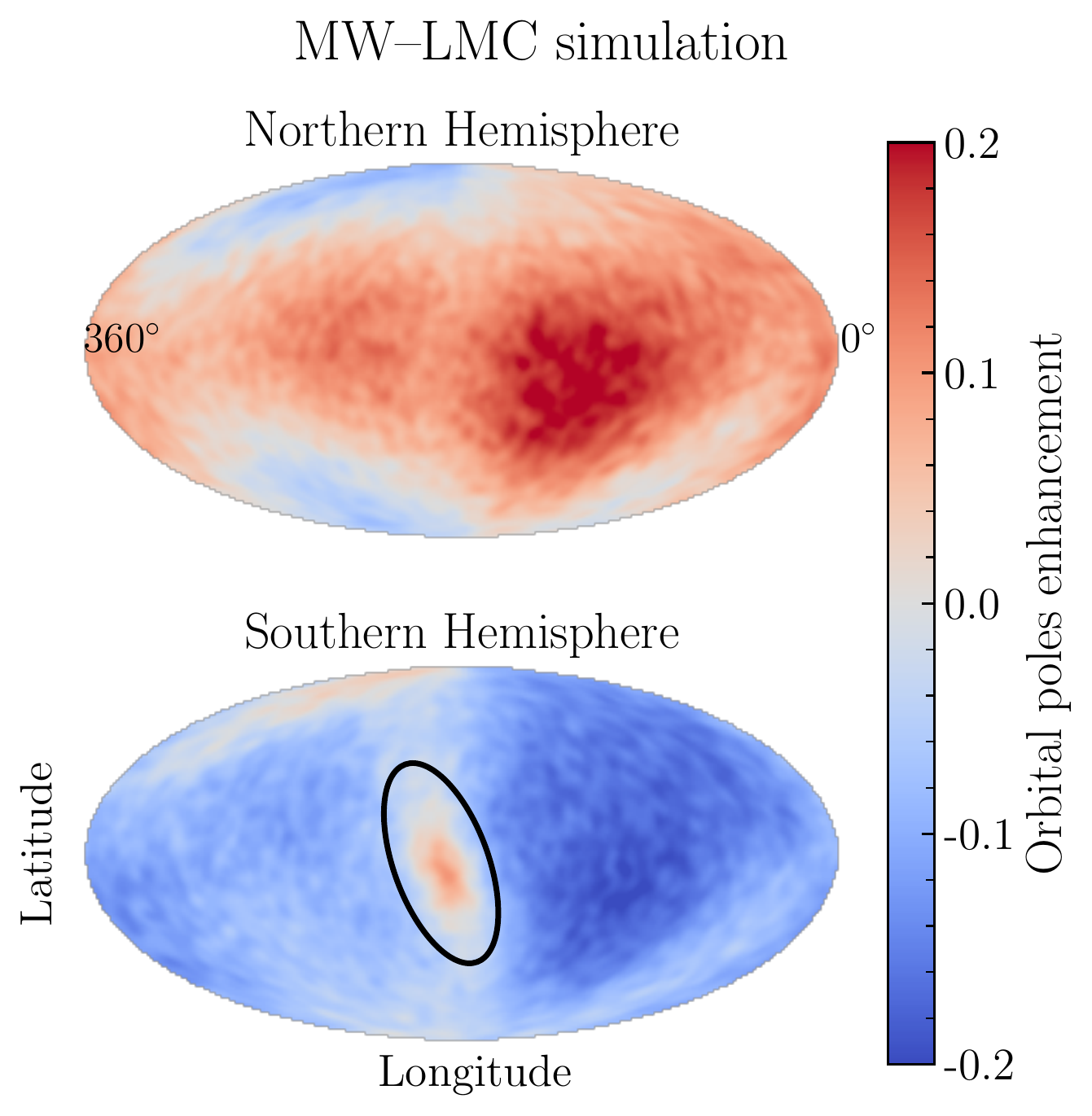}
    \includegraphics[scale=0.41]{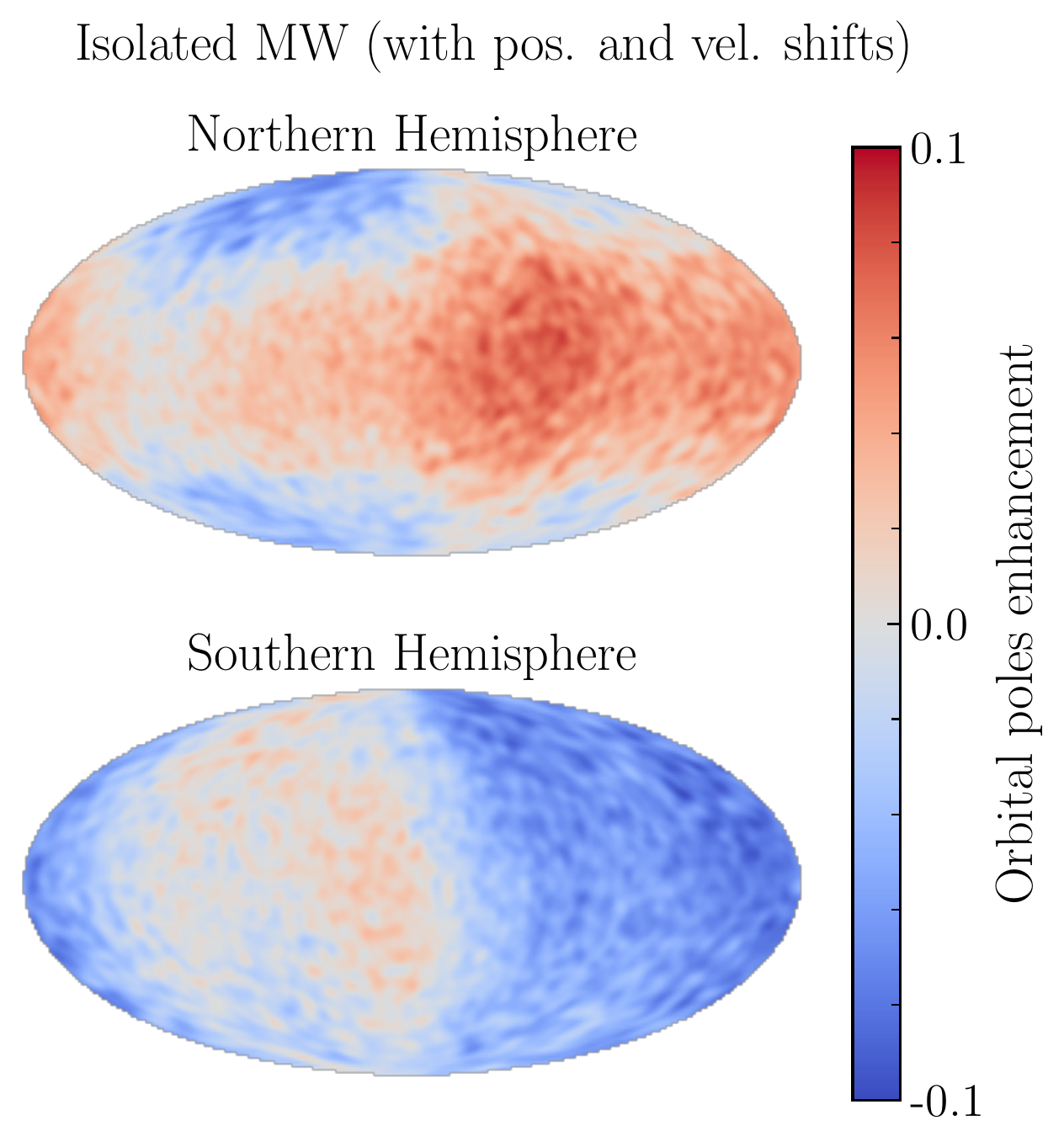}
    \includegraphics[scale=0.43]{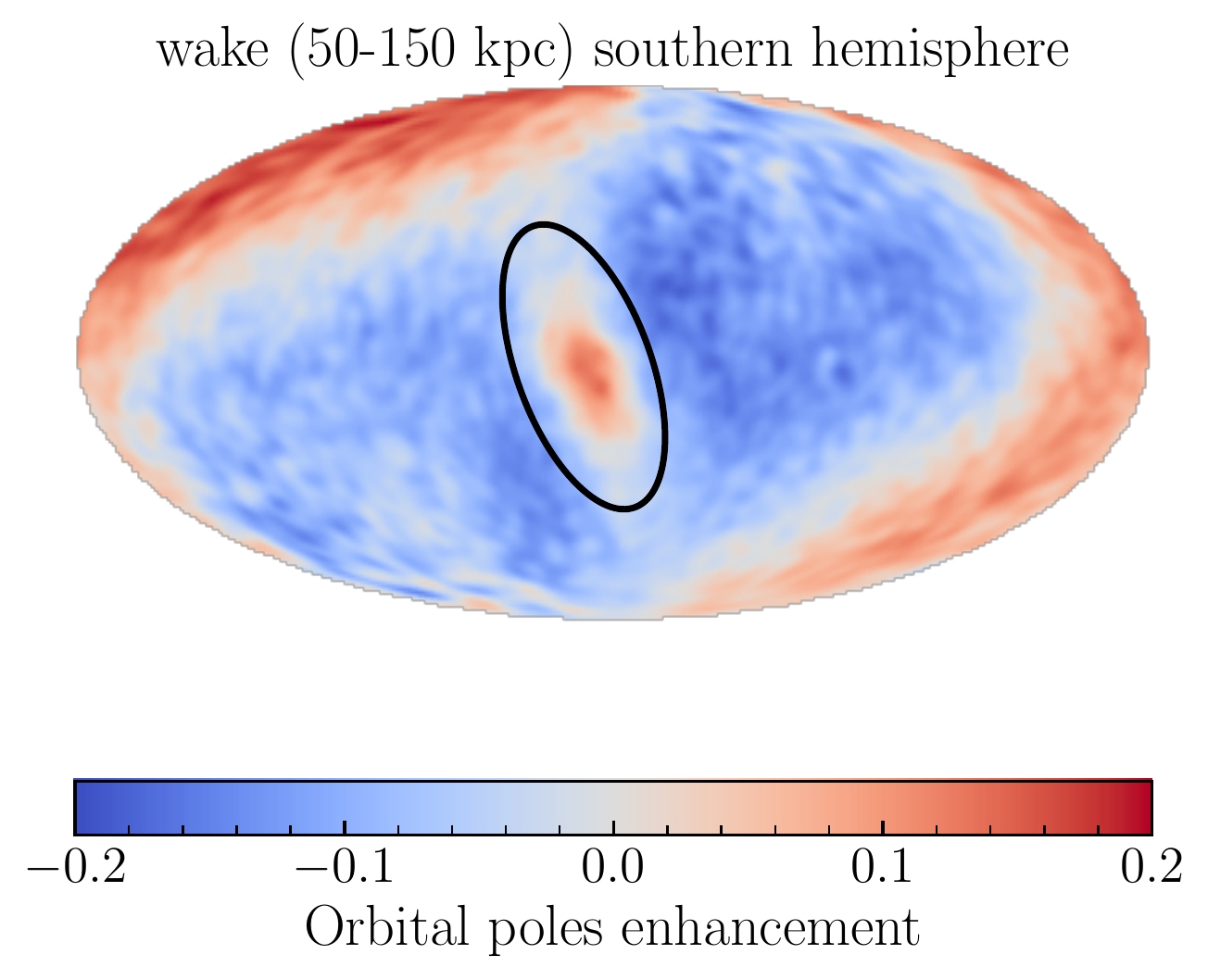}
\caption{Enhancement in the density of orbital poles for the MW DM halo particles selected in the northern and southern hemispheres at distances of 50-150 kpc, relative to particles in an isolated MW halo. \textbf{Left column:} The results for the MW--LMC halo simulation. In the northern hemisphere (top panel) there is an overall enhancement in orbital poles over a large area of the sky. In the southern hemisphere (bottom panel), there is a concentrated overdensity at Lon=180$^{\circ}$ as shown by the black ellipse. \textbf{Middle column:} The location of orbital poles of particles in the isolated MW halo shifted in both positions and velocities to account for the COM displacement (\S \ref{sec:com_displacement}) and reflex motion (\S \ref{sec:reflex_motion}). Qualitatively, the features of the MW--LMC halo can be reproduced, particularly in the north, but less well in the south.  
\textbf{Right panel:} The enhancement of orbital poles using particles selected within the volume of the dynamical friction wake in the MW--LMC simulation relative to the particles in that same volume in the isolated MW. 
Particles were selected within a cylinder of radius 60 kpc, centered on the LMC's past orbit in the southern hemisphere. The dynamical friction wake yields an enhancement in orbital poles around Lon=180$^{\circ}$ tilted by 20$^{\circ}$ relative to longitude 180$^{\circ}$ like in the bottom left panel. This demonstrates that the dynamical friction wake induced by the LMC can induce a tight clustering of orbital poles near the orbital pole of the LMC at distances of 50-150 kpc.} 
    \label{fig:north_south_50_150}
\end{figure*}

In previous sections, we studied the effect of the COM displacement and the reflex 
motion on the orbital poles of DM halo particles. We saw that both of these effects could 
reproduce the general sinusoidal patterns seen in orbital pole maps of particles in the MW--LMC simulations, as shown in Figure~\ref{fig:poles_summary}. Here, we explore the enhancement in the clustering caused by the DM dynamical friction wake and the collective response  in the southern hemisphere. 

\subsubsection{The DM dynamical friction wake}
DM particles in the dynamical friction wake induced by the LMC are expected to follow the path of the LMC's orbit. Consequently, those particles should have similar orbital poles to that of the LMC.
The dynamical friction wake is predominantly in the southern hemisphere at distances from 50-150 kpc (G19).
As such, we expect that the all-sky projected distribution of orbital poles of particles in the northern and southern
hemispheres will differ at distances of 50-150 kpc. This is illustrated using particles in the MW--LMC simulation in the left column in Figure~\ref{fig:north_south_50_150}.

In the northern hemisphere (top-left panel) there is an overall enhancement in the poles with a density peak near Lon=$120^{\circ}$ and Lat=$-10^{\circ}$. This is markedly different from the pattern in the south (bottom left). The distribution of orbital poles in the north is caused primarily by the COM displacement and reflex motion. This is illustrated in the top-middle panel, which shows the northern hemisphere of the isolated MW (no LMC) simulation, where shifts in particle positions and velocities were added to mimic the COM displacement and reflex motion, as in Section~\ref{sec:com_displacement} and \ref{sec:reflex_motion}. The overall enhancement in the poles can be reproduced with these shifts in the north. Note, however, that there is a difference in the amplitude by a factor of 2. This reinforces our earlier point that the COM displacement is not well captured by computing the COM in spherical shells.  

In the southern hemisphere, there is an elongated overdensity near Lon=180$^{\circ}$ in the MW--LMC simulation (bottom left panel). This confirms that the concentration of orbital poles near the LMC is driven by changes in the kinematics of particles in the south. As shown in the middle-bottom panel the amplitude of this clustering cannot be reproduced with shifts in positions and velocities alone even though there is a clear enhancement in that region.

To isolate the effect of the dynamical friction wake we select MW DM particles within a twisted cylinder (like a hose) of radius 60 kpc around the LMC's past orbit in the southern hemisphere. This exercise is done using particles in both the isolated MW and the MW--LMC simulation. The right-most panel of Figure~\ref{fig:north_south_50_150} illustrates the relative orbital pole enhancement in the MW--LMC simulation for these particles. We find that the dynamical friction wake does induce a clustering at 
Lon=180$^{\circ}$ and Lat=$0^{\circ}$, highlighted with the black ellipse. This is the same signature observed in the bottom left plot. This result confirms that at distances of 50-150 kpc the dynamical friction wake also imprints a characteristic signal in the density of orbital poles, in addition to the COM displacement and reflex motion.

\subsubsection{The Collective Response }

\begin{figure*}[h]
    \centering
    \includegraphics[scale=0.41]{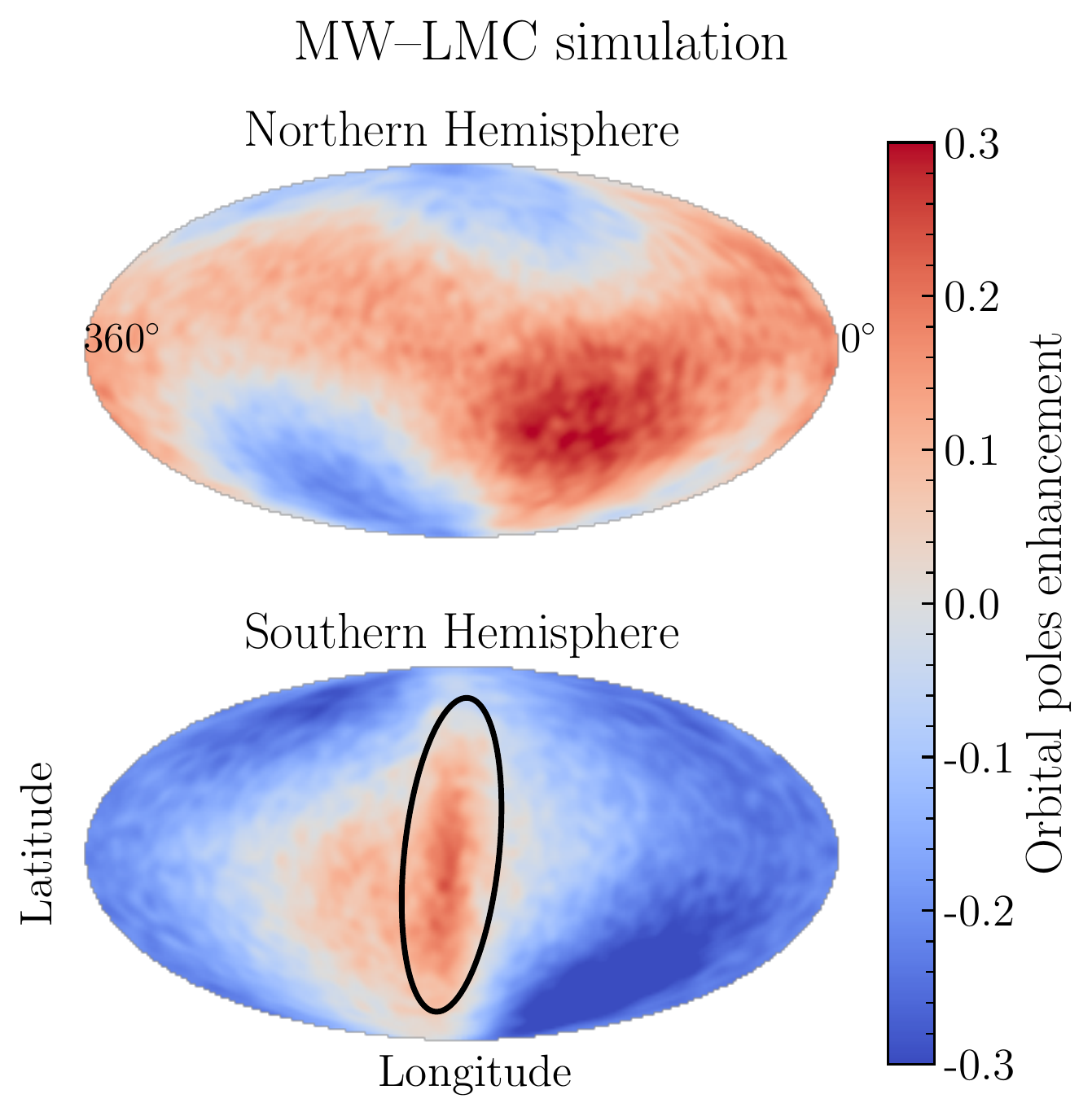}
    \includegraphics[scale=0.41]{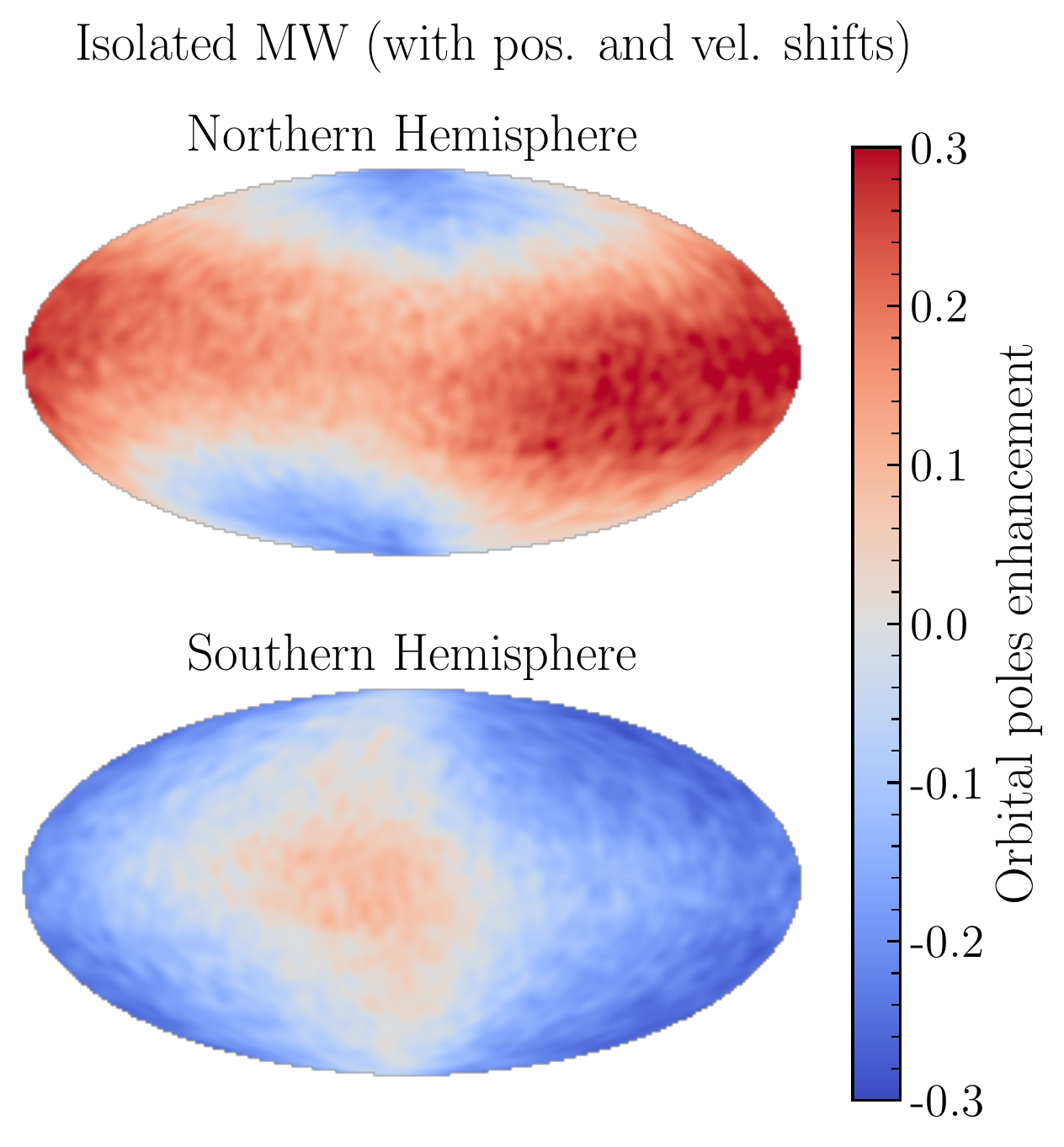}
    \includegraphics[scale=0.43]{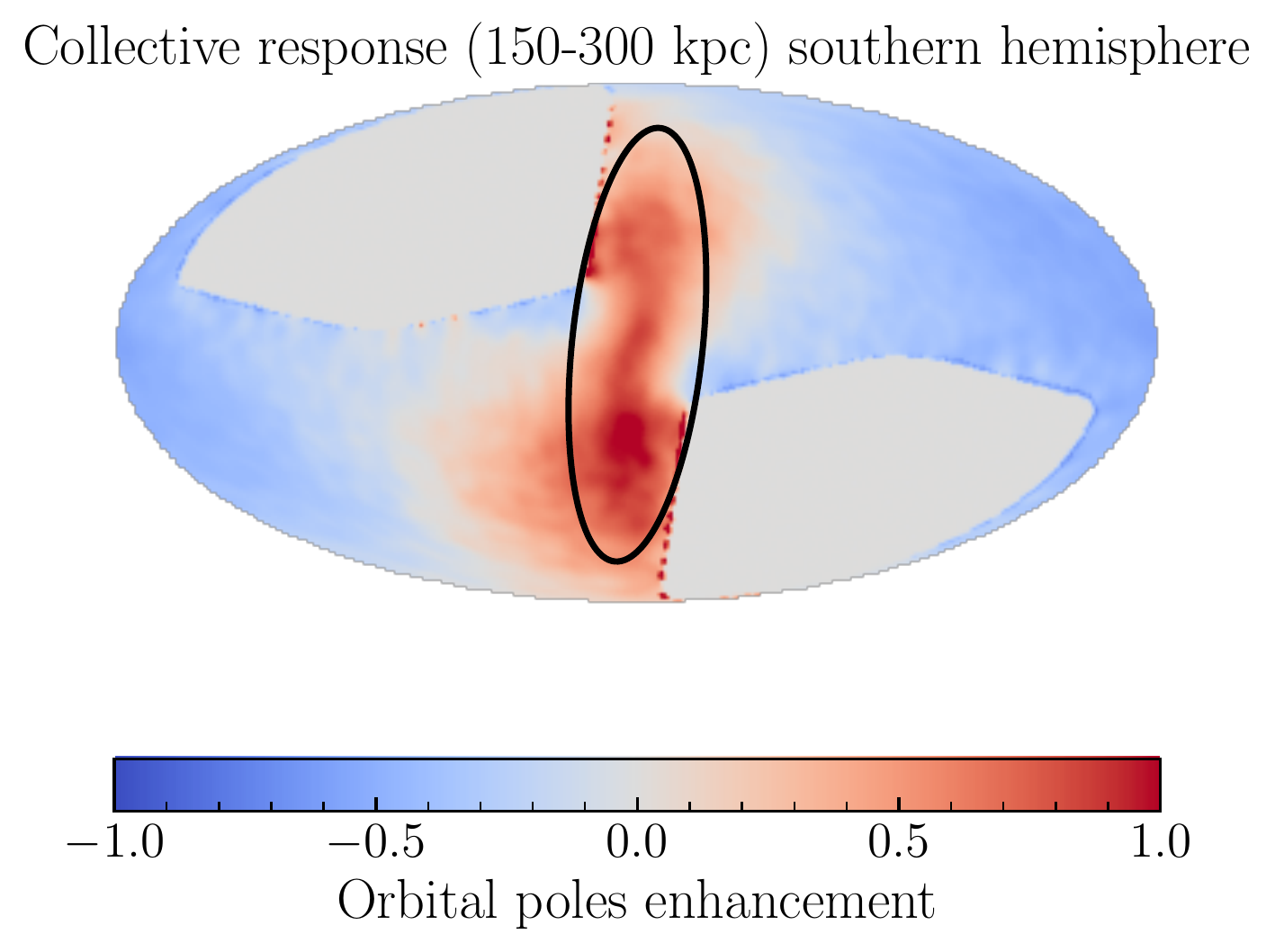}
\caption{Same as Figure~\ref{fig:north_south_50_150} but for MW DM halo particles selected from the MW--LMC simulation at distances of 150-300 kpc. \textbf{Left column:}
In the northern hemisphere (top panel) there is an overall enhancement in orbital poles along a sinusoidal pattern.  However, in the southern hemisphere (bottom panel) the overdensity centered at Lon=180$^{\circ}$ is elongated in a different direction than the particles in the southern hemisphere within 50-150 kpc (see Figure~\ref{fig:north_south_50_150}), as illustrated by the black ellipse. \textbf{Middle panel:} Qualitatively, we can reproduce the orbital pole map of particles in the MW--LMC halo by shifting MW halo particles in the isolated MW simulation in both positions and velocities, but again, the southern hemisphere is not well-reproduced.  
\textbf{Right panel:} The enhancement of orbital poles of particles selected within the volume that encompasses the Southern collective response  (particles at $150<r_{\rm GC}<300$ kpc, $z<0$ kpc, $y<0$ kpc, and $|x|<40 $kpc). Where $x, y, z$ are Cartesian Galactocentric coordinates, where the Sun located at $x=-8.3$ kpc and the northern hemisphere is at $z>0$ kpc. Particles within the collective response  show an enhancement in orbital poles relative to particles in the same volume in the isolated MW simulation. Note that the values of the enhancement are not the same, since the volumes in both cases are different (and therefore some regions are masked in gray due to low particle counts). Note that because the chosen volume is much smaller than a hemisphere, the color scale for this panel should not be directly compared to the other panels. The enhancement is centered at Lon=180$^{\circ}$ and tilted by 6$^{\circ}$ relative to longitude 180$^{\circ}$ and is very similar to that seen in the bottom left panel. This demonstrates that the collective response  impacts the distribution of orbital poles at distances from 150-300 kpc.  }
    \label{fig:north_south_150_300}
\end{figure*}

At distances of 150-300 kpc, the DM dynamical friction wake is not pronounced in the southern hemisphere (G19). However, the collective response  is present (See Figure 2 in G20). As such, we expect the structure in the southern hemisphere at these distances to differ from that illustrated between 50-150 kpc.

Figure~\ref{fig:north_south_150_300} shows the enhancement of orbital poles at distances between 150-300 kpc. In the MW--LMC simulation (left panels), the northern and southern hemispheres again have distinct distributions of orbital poles, as in the case between 50-150 kpc. In the north, the enhancement is along a sinusoid that can be qualitatively explained with the COM displacement and reflex motion (middle top panel). In the southern hemisphere, however, there is a strong enhancement highlighted with the black ellipse. This signature cannot be reproduced with the positions and velocity shifts (bottom middle panel). %Moreover, there is no wake in the south between 150-300 kpc. 

The primary distortion to the MW halo at distances  % overdensity that is at distances 
of 150-300 kpc is the collective response , which is a large overdensity in the MW's DM halo that encompasses the majority of the northern hemisphere (G20). However, in the southern hemisphere, the collective response  is confined within a smaller volume that aligns with the $y-z$ Galactocentric plane. 

The collective response  has a complex shape that cannot be captured by simple shifts in the COM position. As such, we study this structure by selecting particles that we know are located within the overdensity in the southern hemisphere. %in the south is part of the collective response . 
We define particles to be in the Southern collective response  if they reside in the following volume in the southern hemisphere: $150<r_{\rm GC}<300$ kpc, $z<0$ kpc, $y<0$ kpc, and $|x|<40$ kpc. Where $x, y, z$ are Cartesian Galactocentric coordinates.

The right panel of Figure~\ref{fig:north_south_150_300} shows the enhancement of the orbital poles for particles in this volume.

Particles located within Southern collective response  exhibit a strong enhancement in the density of orbital poles, highlighted with the black ellipse. The enhancement is in the same region as the one observed in the bottom left panel of Figure~\ref{fig:north_south_150_300}. The same pattern is not seen in the Northern Hemisphere (which is dominated by the Northern collective response ). The difference is likely a result of the strong coincidence of the Southern collective response  with the $y-z$ Galactocentric plane, v.s. the large angular coverage of the Northern collective response  (see the next section). 

This exercise demonstrates that the collective response  is not well captured by a simple COM displacement plus the reflex motion, and that particles in the Southern collective response  leave an important imprint in the distribution of orbital poles on the sky.

%%%%%%%%%%%%%%%%%%%%%%

\section{The Clustering of Orbital Poles Induced by Selecting Planar Configurations}\label{sec:bias}

In a homogeneous spherical system of particles in equilibrium, the density of orbital poles of those particles is also homogeneous. 
This is shown in the left panel of Figure~\ref{fig:MWLMCsims}, where particles are selected within a spherical shell centered on 
the Galactic center in the isolated MW simulation. However, if one selects particles in a smaller angular region of the spherical halo, the resulting density of orbital
poles would have a particular, non-homogeneous pattern on the sky. In the case of particles selected within a planar configuration, 
the resulting density of orbital poles will cluster in the direction of the plane's minor axis.   

\begin{figure}[h]
\centering
\includegraphics[scale=0.5]{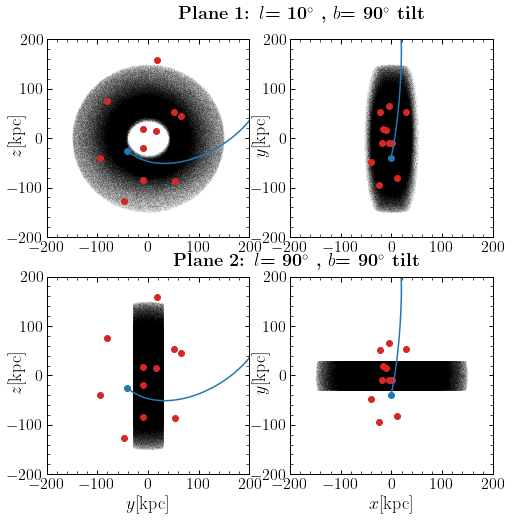}
\caption{Distribution of DM particles (black dots) selected within two planar configurations in the simulation, projected in $y-z$ (left) and $x-z$ (right) Galactocentric planes. Each plane has a thickness of 30 kpc and extends from 50-150 kpc radially from the Galactic center. Plane 1 (top row) is defined as the orbital plane of the LMC, which is also roughly the $y-z$ Galactocentric plane. The blue solid line indicates the past orbit of the LMC. Plane 2 is perpendicular to the LMC's orbit, similar to 
the orbit of the Sagittarius dwarf galaxy. Satellite galaxies that are associated with the VPOS are superposed in the simulations shown in red circles. VPOS satellites currently reside within Plane 1.}
\label{fig:planes}
\end{figure}

We illustrate this clustering of orbital poles by first selecting halo particles in the isolated MW simulation that are confined within 
two planes centered on the Galactic center (see Figure~\ref{fig:planes}). Plane 1 is defined as the orbital plane of the LMC, which is roughly the 
Galactocentric $y-z$ plane. This is also the plane in which the majority of the VPOS satellites reside. Plane 2 is perpendicular to Plane 1, similar to 
the orbit of the Sagittarius dwarf galaxy.

\begin{figure}[h]
\centering
\includegraphics[scale=0.5]{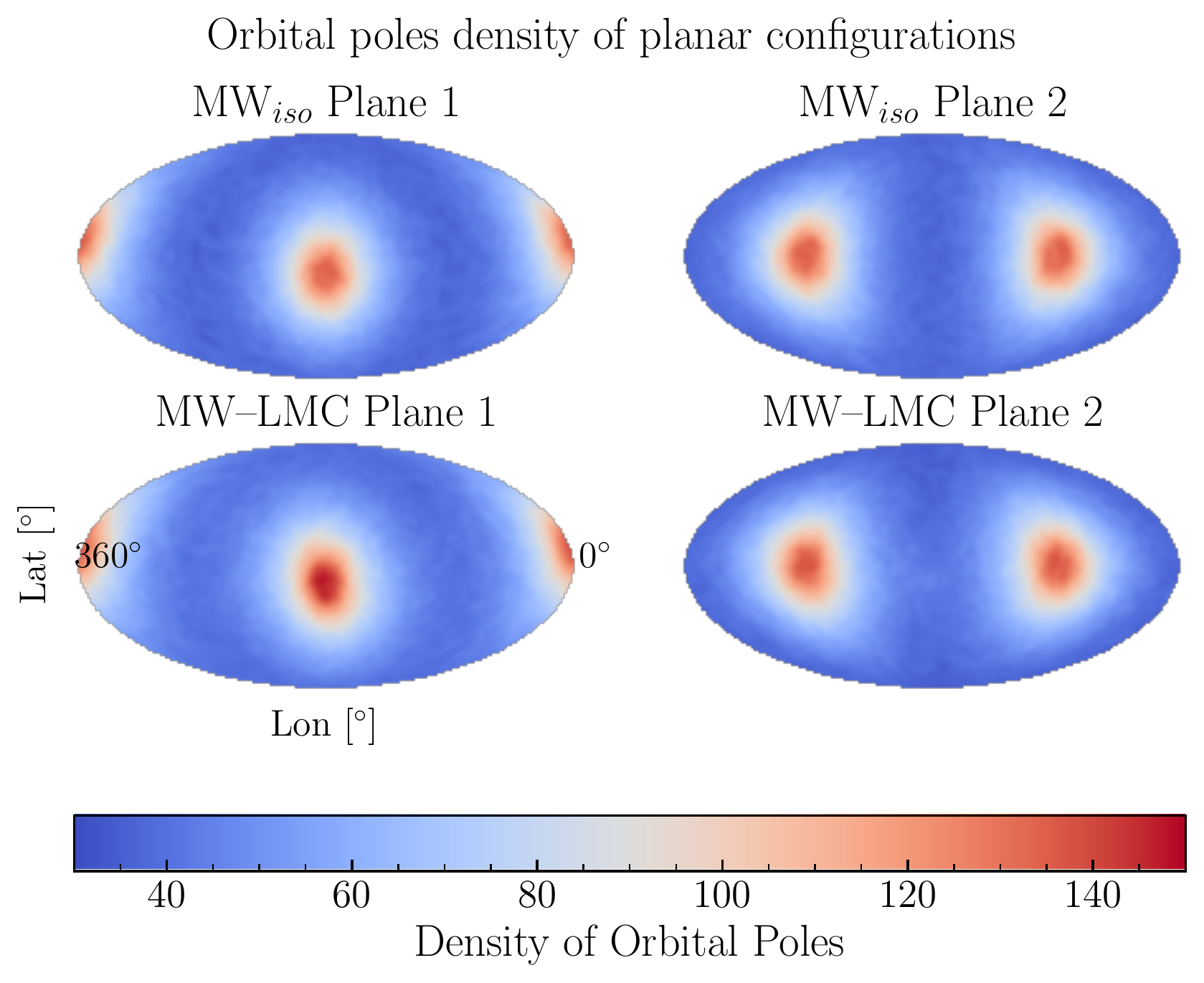}
\caption{All-sky density map of orbital poles of DM particles selected within the two planes illustrated in Figure~\ref{fig:planes}. For both planes the distribution of orbital poles is aligned with the minor axis of the planes. For Plane 1 the clustering is close to the LMC's orbital pole (Lat$=0^{\circ}$, Lon$=180^{\circ}$) and the counter-rotating regions (Lat$=0^{\circ}$, Lon=$360^{\circ}$). In the MW--LMC simulations, the clustering of poles occurs in the same regions as in the isolated case, but for Plane 1 there is an enhancement in the clustering of orbital poles by $ \delta \rho_{iso} \approx\approx 20\%$.
%(see Figure~\ref{fig:enhacement_planes}. 
Plane 2 also exhibits a $\delta \rho_{iso} \approx 20\%$ enhancement near the LMC's orbital pole, barely seen since it is located in an underdensity of orbital poles.
%is not noticeable. 
This exercise illustrates that particles with isotropic velocities selected within planar configurations will have an asymmetric distribution of orbital poles, where the density of poles will concentrate near the minor axis of the plane. The inclusion of the LMC appears to enhance this effect in its own orbital plane.}
\label{fig:bias_planes}
\end{figure}

\begin{figure}[h]
    \centering
    \includegraphics[scale=0.5]{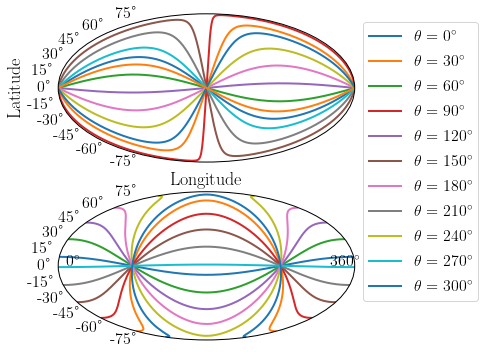}
    \caption{Orbital poles of particles in Planes 1 (top panel) and 2 (bottom panel) evenly distributed in polar angles. For each particle the position is fixed and the orbital poles are computed with 1000 isotropic velocity vectors. The resulting orbital poles for each position form a sinusoidal pattern as shown in each color. As the position along each plane changes the sinusoid changes amplitude but the period remains constant. The intersection of all the sinusoids create the overdensities observed in Figure~\ref{fig:bias_planes}. This illustrates the bias in orbital poles when selecting planar configurations.}
    \label{fig:poles_bias}
\end{figure}

In the top row of Figure~\ref{fig:bias_planes}, the density of orbital poles of particles in the isolated MW simulation that are located in these two planes are plotted in Galactic coordinates. In each case, the orbital poles are clustered along the minor axis of each plane (Lon=$180^{\circ}$, Lat=$0^{\circ}$). There are two overdensities that have an angular separation of 180$^{\circ}$ corresponding to co-rotation and counter-rotation. 
As shown in Section \ref{subsec:reflex}, when we introduced a displacement in the velocity vector corresponding to the reflex of the COM motion of the MW's disk, we found a resulting sinuosoidal pattern in the distribution of orbital poles. Similarly, if we select particles located in different sections of each plane based on their polar angle, and thereby further confine the position of the particles, we would get a series of sinusoids, as illustrated in Figure~\ref{fig:poles_bias}. The intersection of the sinusoids occurs every 180$^{\circ}$, creating the corresponding clustering of the orbital poles seen in Figure~\ref{fig:bias_planes}.

This effect was described in \citet{Johnston96, Mateu11, Mateu17} and has since been used to discover and characterize stellar streams and substructure 
in the stellar halo \citep[e.g:][]{Mateu18, Ramos20}. The clustering of orbital poles for satellites spatially selected to reside in a 
plane is an expectation of this bias. {\color{black}In the particular case of the VPOS, the effect of the selection function of SDSS was discussed in \cite{Pawlowski16}, where the VPOS was found to be 5$\sigma$ more significant than an isotropic distribution of satellites. As shown in this section, that planar configuration will induce a clustering in satellite orbital poles.}

The next question is whether the clustering of poles is more pronounced with the inclusion of the 
LMC than expected for a purely isotropic distribution.
We repeat this same exercise by selecting the same planes, but using the simulation of the MW that includes the LMC. The resulting density of orbital poles for DM particles in Plane 1 and Plane 2 are illustrated in the bottom row of Figure~\ref{fig:bias_planes}. Again, we see two overdensities in the projected distribution of orbital poles on the sky and at the same locations. However, the LMC enhances the amplitude of the clustering by $\delta \rho_{iso}\approx$20\%, centered on its orbital pole, for the reasons described in the previous sections.

We conclude that the orbital poles of objects orbiting the MW that are selected within a planar configuration will exhibit 
orbital poles that cluster around both the projected location of the normal to the selected plane, and 180$^{\circ}$ from that normal. The inclusion of the LMC augments this effect in the $y-z$ Galactocentric plane, resulting 
in a higher concentration of orbital poles near the VPOS than expected in the purely isotropic case, especially in the co-rotating region.

\section{Discussion}
\label{sec:discussion}

Here we discuss the implications of our results for halo tracers (Section \ref{sec:tracers}), the time dependence of the simulated clustering of orbital poles (Section \ref{sec:VPOStime}), and we further discuss our findings in the context of planes of satellites around other galaxies (Section \ref{sec:externalplanes}) and in cosmological simulations (Section \ref{sec:cosm_planes}).

\subsection{Implications for Halo Tracers}
\label{sec:tracers}

\begin{figure}
    \centering
    \includegraphics[width=\columnwidth]{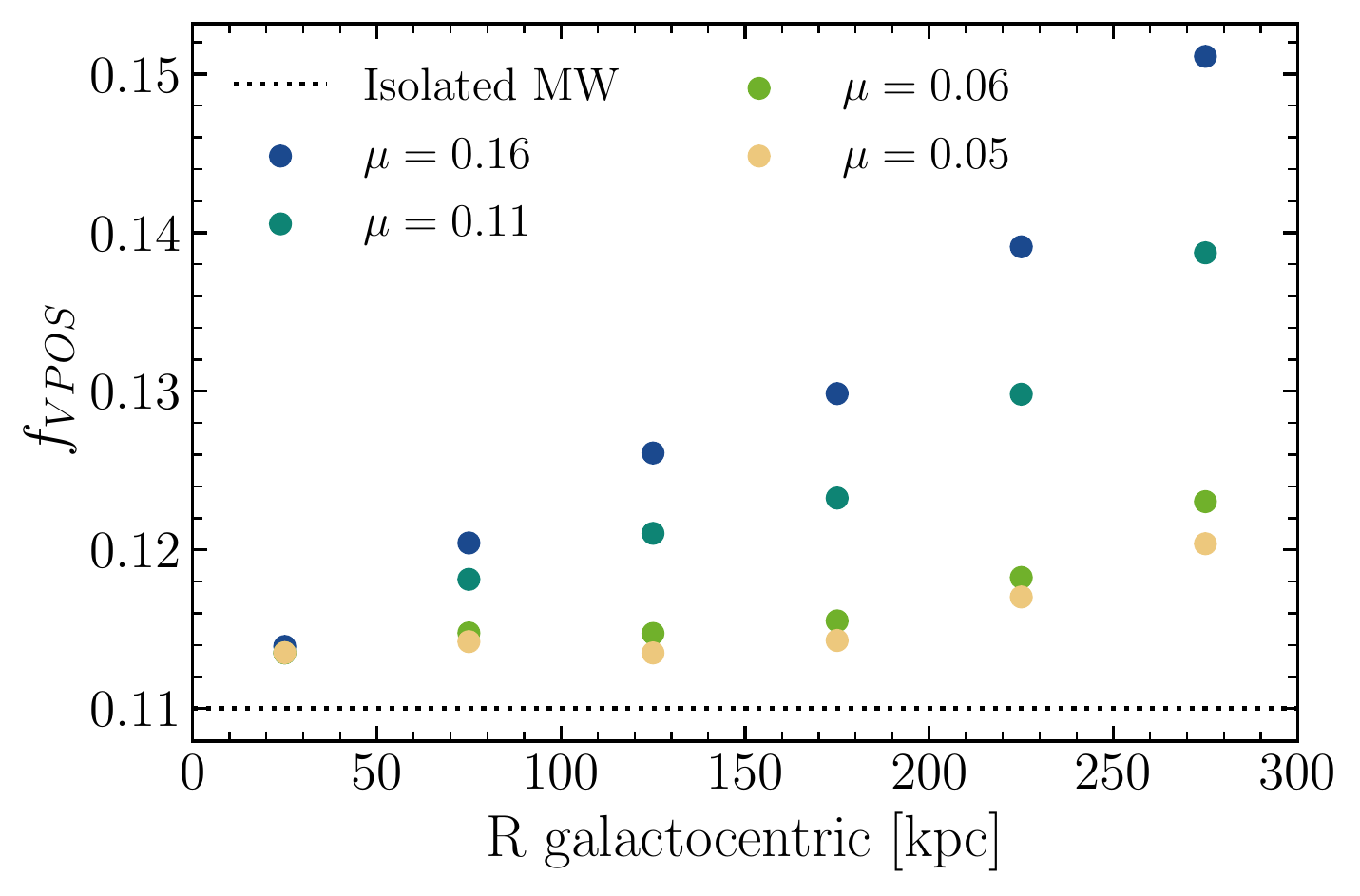}
    \caption{Fraction of particles that have orbital poles within a 36.87 degree area, centered on the orbital pole of the simulated LMC ($f_{VPOS}$) as a function of Galactocentric radius. This area roughly corresponds to the VPOS (counter-orbiting particles are not included). $f_{VPOS}$ is defined as the ratio of particles with orbital poles in the VPOS over all particles in a shell of 50 kpc thickness ($f_{VPOS} = \frac{N part. VPOS }{N part. shell}$). %We use the same VPOS definition as in Figure~\ref{fig:MWLMCsims}, but centered on the LMC's orbital pole in each simulations. 
    We exclude particles with latitudes smaller than $20^{\circ}$ to avoid regions that are harder to observe due to the MW's disk. The black dotted line shows $f_{VPOS}$ for the isolated MW simulation, indicating that for a spherically symmetric system in equilibrium, $\approx$10\% of particles will have orbital poles in the VPOS (see Figure~\ref{fig:poles_bias}). $f_{VPOS}$ is computed for 4 simulations using different MW--LMC mass ratios ($\mu = M_{lmc}/M_{mw}$). The fiducial model corresponds to $\mu=0.11$. We find that $f_{VPOS}$ increases as a function of both Galactocentric radius and $\mu$, peaking at 15\%.} 
    \label{fig:fvpos}
\end{figure}

In this work we have analyzed the orbital poles of DM particles in a MW--LMC simulation compared to an
isolated MW halo simulation. This has allowed us to study a generic mechanism that
induces clustering in the distribution of orbital poles due to the perturbations of a massive
satellite. We expect that the same mechanism would affect halo tracers, such as
globular clusters, satellite galaxies, stellar streams and halo stars. As such,
in this new scenario,
the orbital poles of such tracers (beyond 30 kpc)
would also be biased towards the LMC's orbital pole. 
We can estimate the fraction of halo tracers that may be impacted by this effect, assuming they are initially in an isotropic distribution within the MW halo. 
We first take all DM particles in spherical shells, 50 kpc in thickness, centered at different Galactocentric radii.  
In Figure~\ref{fig:fvpos} we plot the fraction of those DM particles ($f_{VPOS}$) that have orbital poles within a 36.78 degree area, centered on the orbital pole of the simulated LMC (i.e. approximately the VPOS region).

This exercise is computed using 4 different MW--LMC simulations from G19 (simulations 5-8; anisotropic MW dispersion profile), where the LMC mass is assumed to be $8, 10, 18, 25 \times10^{10}$ M$_\odot$. The corresponding mass ratio between the LMC/MW is $\mu = 0.05,0.06, 0.11, 0.16$. Our fiducial simulation is $\mu = 0.11$.

We find that $f_{VPOS}$ increases as a function of the LMC-MW mass ratio ($\mu)$. For the most massive LMC models, $f_{VPOS}$ reaches up to 14-15\% in the outer halo ($> 200$ kpc), which is consistent with the 20-25\% enhancement relative to the isolated case (dashed lines) that we found in Figure~\ref{fig:MWLMC_distance}. Higher LMC infall masses have been suggested in the literature based on cosmological expectations \citep[as high as $3.4 \times 10^{11}$M$_\odot$][]{shao18, besla19}, which can help to augment this fraction.   

\textit{We expect that the orbital poles of halo tracers, such as satellite galaxies, stars, globular clusters, and streams, that
pass through volumes of the halo that encompass the DM dynamical friction wake or
collective response  will be similarly affected}. Testing this assertion for
specific satellites will require integrating orbits in a time-evolving MW--LMC
potentials that capture the DM dynamical friction wake and collective response;
this will be the subject of future work.

We caution against using the quoted fractions as a direct comparison to the currently observed fraction of satellites with orbital poles in the VPOS, as:

\begin{enumerate}
   \item The full observational census of MW satellites is not yet complete and membership to the VPOS depends on astrometric measurements with large uncertainties. The latter also impacts the identification of LMC satellites, whose kinematics would not be influenced by the effects described in this paper.

\item We have also assumed initial conditions for the MW as a spherical halo in equilibrium, and, as such, these idealized simulations do not account for the cosmological assembly history of the MW. In particular, satellite accretion is not expected to be isotropic and we have not followed the infall of the LMC from large distances. It is not yet clear if an accreted stellar halo of the MW would be affected in the same way as a spherical DM halo. For example, the amplitude of the MW's stellar wake \citep{conroy21} is larger by a factor of 2 with respect to the DM wake in the simulations.
\end{enumerate}

Instead, these results should be considered a lower bound on the possible impact of the LMC on the kinematics of halo tracers. Observations of the VPOS clustering motivates future simulations that take these factors into account. Confirmation that the LMC influences the formation of the VPOS would be of great importance to disentangle the accretion history of the Galaxy and constrain alternative models of DM and gravity.

\subsection{The Time Dependence of the Clustering of Orbital Poles}
\label{sec:VPOStime}

\begin{figure}
    \centering
    \includegraphics[scale=0.9]{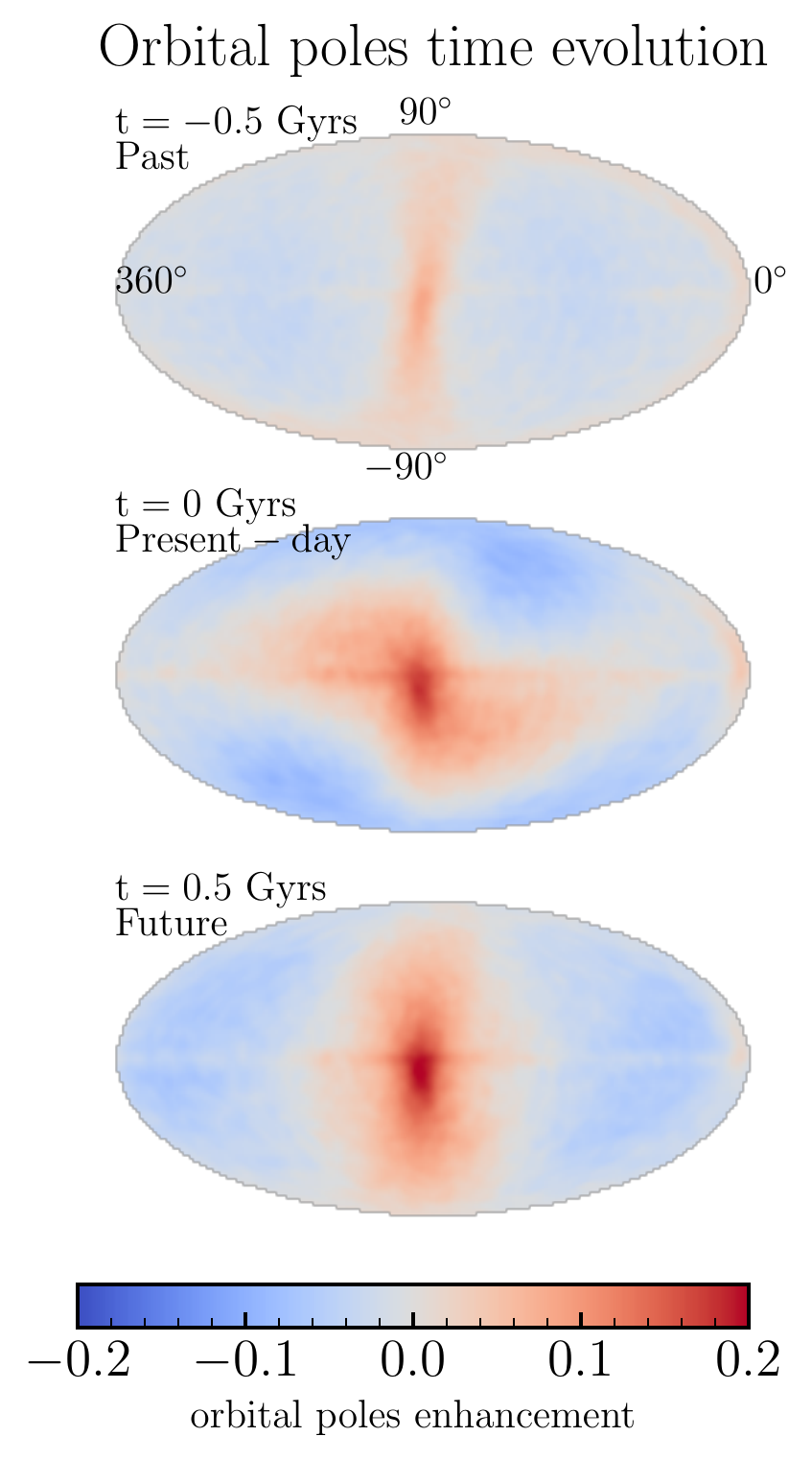}
    \caption{All-sky Mollweide projections depicting the temporal evolution of the density of orbital poles in the MW--LMC simulation, relative to the isolated MW model (left panel of Figure~\ref{fig:MWLMCsims}).
    For all the maps the distance range is from 0-400 kpc. The middle panel shows the present-day results (similar to the right panel of Figure~\ref{fig:MWLMCsims}).
    The top panel shows the density of orbital 
    poles for particles in the MW's halo 0.5 Gyrs ago, when the LMC made its first pericentric approach.  The bottom panel shows the orbital distribution 0.5 Gyrs in the future. The enhancement of orbital poles ($\delta \rho_{iso}  \geq 15\%$) clustered around the LMC still exists $\pm$0.5 Gyr from now. The changes to our reference frame last at least $\approx$1 Gyr. New simulations are required to assess the longevity of this structure beyond this timescale.}
    \label{fig:VPOSt}
\end{figure}

In this section, we explore the longevity of the simulated clustering of orbital poles induced by the LMC. Figure~\ref{fig:VPOSt} shows the temporal evolution of orbital poles for all MW halo particles in the MW--LMC simulation.

The clustering of orbital poles is noticeable starting 0.5 Gyr ago and persists 0.5 Gyr into the future. However, additional simulations are needed to quantify how much longer the clustering will persist. We can state only that this clustering lasts for at least 1 Gyr. This timescale is consistent with that reported for the longevity of planes found in cosmological zoom-in simulations of MW analogs that host LMC analogs \citep[][see Section~\ref{sec:cosm_planes}]{Samuel20}.

\cite{Samuel20} find that the MW-like analogs in the FIRE cosmological simulations that host a massive satellite on their first pericentric passage, are more likely to have longer lasting planes (0.5--1 Gyr) than the MW-mass galaxies without an LMC analog. The latter unanimously have plane lifetimes $<$ 500 Myr. It should be noted that plane lifetimes are characterized in \cite{Samuel20} as halos exhibiting the same planar properties (i.e. orbital dispersion, axis ratio, RMS height) for consecutive simulation snapshots, whereas our findings are based on an orbital pole enhancement in a MW halo including the presence of the LMC as viewed from a non-inertial reference frame. Our findings support the idea that the LMC can promote the strength and longevity of satellite planes. 

If the clustering of orbital poles is transient, this could potentially explain why planar configurations of satellites are rare in cosmological simulations. Such induced alignments may only be observable during pericentric passages of massive satellites and last for only $\approx$1 Gyr. Furthermore, the satellite needs to be both sufficiently massive and have a sufficiently close approach to induce a substantive change in the relative motion and spatial location of our reference frame relative to objects in orbit within the halo. These results suggest that the MW is currently in a unique time in its evolution.

\subsection{Implications for Extragalactic Planes}
\label{sec:externalplanes}

Planes of satellites have been proposed to exist around
other MW-mass galaxies in the Local Volume, including Andromeda (M31), Centaurus A (CenA), and other galaxies of approximately the same mass \citep{Zaritsky99, Ibata2013, Conn2013, Shaya2013, Muller2018, Libeskind2019, Muller20}. Many of the known satellites orbiting M31 and Cen A are posited to 
either co-rotate or counter-rotate within one or more thin orbital planes, 
similar to satellites in the VPOS. Although 3-D velocity measurements are not yet available to confirm kinematic coherency for Cen A, data for some M31 satellites is available. Radial velocities of satellites in the Great Plane of Andromeda (GPoA) suggest coherent rotation and recent HST proper motion measurements for NGC 147 and NGC 185 support their association with the GPoA
\citep{Sohn20}. Forthcoming proper motion measurements with JWST and HST (e.g. GO-15902, PI Dan Weisz and GO-16273 PI Tony Sohn) will provide additional 3-D velocities for all remaining M31 satellites.

While our proposed mechanism may explain the tight correlation in the orbital poles of a subset of MW satellites with that of the LMC, it remains unclear whether this mechanism is relevant for other, extragalactic planes, such as the GPoA. 

A logical analogy to the MW--LMC system is M31's most massive surviving satellite galaxy, M33, whose infall mass is likely similar to, or higher than, that of the LMC. M33 exhibits a pronounced stellar and gaseous warp, which could arise if M33 had a close ($\approx$50 kpc) approach to M31 \citep[e.g.][]{Mcconnachie09,Putman09}. However, recent HST, VLBA, and Gaia proper motion measurements of M31 result in a relative velocity vector of M33 with respect to M31 that is incompatible with such a close approach \citep{Patel17a, Vdm2019, Sohn2012, Vdm12, Brunthaler05}. Using the tangential velocity derived from proper motions of the M31-M33 system, less than 1\% of M33 satellites ($M_{halo}=2.5\times 10^{11}\, M_{\odot}$) in a high mass M31 model ($2 \times 10^{12}\, M_{\odot}$) have infall times in the last 6 Gyr, a recent pericentric passage ($t < 3$ Gyr ago), and a distance at pericenter of 100 kpc or less \citep{Patel17a}. These measurements and orbital analyses limit the plausibility that M33 induced the GPoA through distortions to M31's halo after a pericentric approach. 
 
We do not claim that the scenario we have presented to explain the clustering of orbital poles is the {\it only} mechanism through which planes can form. M31 has experienced a much more active assembly history than the MW \citep[e.g.][]{Hammer2018,DSouza18, McConnachie2018}. Such past accretion events must impact the structure of M31's halo out to large distances and could promote the formation of planes. The argument we have made here is that the orbital poles of satellites will be biased towards the orbital plane of a massive satellite. However, we do not know how long-lived such structures may be. 

For example, we expect that the clustering of satellite orbital poles may be boosted after each close approach of a satellite to its host. Here, we have only studied the evolution of the resulting plane after the first pericentric approach of the LMC, and thus cannot comment on how long planes may persist after the final coalescence of a massive satellite (or major merger) or for massive satellites in different orbital configurations.

\subsection{Planes in Cosmological Simulations}
\label{sec:cosm_planes}

As proper motions have confirmed the tight spatial and kinematic correlation between the VPOS satellites, several studies have analyzed mid and large volumes, cosmological simulations to determine whether there is a statistically significant sample of host-satellite systems exhibiting similar planar configurations \citep{Libeskind2009, Deason11a, Pawlowski2014a, Cautun15, Pawlowski2018, forero-romero18, Pawlowski19, shao19, Lovell21}. However, only 1--2.5\% of systems in the studied cosmological simulations are able to match the alignment of orbits for between six and eight satellites. Further requiring that these satellites reside within a highly flattened plane, as is the case for the observed VPOS, reduces these statistics to only 0.3\% of systems in large volume simulations. This approach is ultimately limited by the resolution of such large volume simulations where individual DM particles can have masses in excess of $8 \times 10^6 \ M_{\odot}$. Subsequently, several groups have turned to suites of cosmological zoom-in simulations to overcome this resolution barrier, yet these studies have also found that only $0.2\%$ of MW-mass halos exhibit both orbital alignment and spatial flattening \citep{Pawlowski2014b, forero-romero18}.

The effect in the stability of planes in time-dependent potentials of triaxial halos from cosmological simulations was studied in \cite{sanders20}. Orbital poles were not dispersed in the time-dependent potentials for more than $6^{\circ}$. Planes were found to be more stable when they were aligned with the halo's major axis. Finally, the properties of the planes seem to be more affected by late-time accretion events, as also shown in \cite{shao19}. These findings strengthen the idea that the LMC is playing a major role in the formation of the VPOS. 

In this study we showed that the LMC can bias the orbital poles of objects towards its orbital plane. More generically, this scenario should occur after the first pericentric approach of any massive satellite about a more massive host. This bias should strengthen the correlation in phase-space properties of objects that already exist in broader planes, like those predicted in CDM models.   

In the context of the results presented here, an additional challenge for cosmological studies is the selection criteria used to identify MW-like halos. Our study motivates looking specifically at MW-like halos that also host a representative analog of the LMC. \citet{Patel17a} found that about 24\% of MW/M31 analogs were identified to host mass analogs of the LMC/M33 at $z=0$ in the Illustris simulations \citep[see also][]{Busha11, Boylan-Kolchin11, Evans15, santos20b}. However, this statistic reflects a selection based on only the mass of the host-satellite systems. 
When accounting for the close position of the LMC today ($\approx$50 kpc) and a recent infall ($t < 4$ Gyr ago) into the halo of the MW, in addition to mass ratio, only about 6\% of MW mass analogs host an LMC mass companion with these properties \citep[see also][]{Boylan-Kolchin11}. Thus, it is not surprising that both large-volume and zoom-in cosmological simulations
find very few systems matching the properties of the VPOS today, especially when an LMC analog is not present \citep[see also][]{Samuel20, Santos20}. Studies like those of \cite{Samuel20, Santos20} are ideal testing grounds for the work presented in this paper.

The relatively short timescales (at least 1 Gyr) over which the clustering of orbital poles after the infall of massive satellites may persist (Section \ref{sec:VPOStime}), provides yet another reason that studies of cosmological simulations have not recovered a statistically significant sample of VPOS-like satellite systems. Furthermore, the temporal resolution of simulations must be sufficient to discern changes on $\approx 1$ Gyr timescale over which the signature of the VPOS first forms and then disperses. To investigate the nature of planes and their connection to massive satellites on first infall in a statistically significant way, halos must be selected across cosmic time, not only at $z=0$.

{\color{black}

The analysis we presented is solely based on simulations with spherical MW halos. However, in cosmological simulations, halos are triaxial especially in the outskirts (beyond $\approx 20 kpc$ in MW-like galaxies \citep{Valluri13}), and subsequently  the population of orbits is different than in the spherical case (i.e., box orbits, short and long axis tubes) \citep[e.g.,][]{Valluri10, Bryan12, Valluri13, Zhu17}. Orbital poles clustering in such triaxial systems might be different in both amplitude and shape than in the spherical case (i.e different sinusoidal pattern). However, we anticipate that an LMC on first infall will also affect a triaxial halo by inducing a wake and barycenter motion -- two necessary ingredients to induce orbital poles clustering. Quantifying the clustering of orbital poles in such triaxial, cosmological halos will be the subject of future work}

A critical benefit of the model presented here is that the combination of the motion of the observer's reference frame with respect to the outer halo (observed as reflex motion), the displacement of the halo's central density, and the dynamical friction wake, impacts multiple satellites across the entire sky. In contrast, a cosmological group infall scenario would only include a limited number of VPOS members \citep{Patel20, Pawlowski2020, santos20b}. The mechanism presented in this study may provide a natural pathway to form a spatially extended, co-rotating and counter-rotating, planar system of satellites that is consistent with the CDM framework.

This proposed mechanism is not unique to CDM. The changes to the observer's reference frame owing to a close pericentric passage of a massive satellite should occur in SIDM, FDM, WDM models and even in alternative gravity models. It is, however, unclear if, e.g. the mass ratio between satellite and host, would be sufficient in these alternative models to induce the necessary changes to our reference frame to produce a similar clustering of orbital poles.

%%%%%%%%%%%%%%%%%%%%%%%%%%%%
\section{Summary and Conclusions}\label{sec:conclusions}

In this study we have demonstrated that the orbital poles of simulated MW DM halo particles will cluster owing to the recent close passage of the Large Magellanic Cloud (LMC). We utilize a tailored N--body simulation of the LMC's passage about the MW from G19 (their Model 7, LMC halo mass at infall of $1.8\times10^{11}$ M$_\odot$) to illustrate that the projected distribution of orbital poles of MW DM particles on the sky will be enhanced along a sinusoidal pattern that clusters near the orbital pole of the LMC and VPOS (Figure~\ref{fig:MWLMCsims}). 

The density of orbital poles is enhanced near the LMC by a factor $\delta \rho_{max}\approx$30\% (50\%) with respect to underdense regions, and $\delta \rho_{iso}\approx $15\% (25\%) relative to the isolated MW simulation (no LMC) over 50-150 kpc (150-300 kpc). The enhancement starts at 50 kpc, where the amplitude of this enhancement increases as a function of distance (see Figure~\ref{fig:MWLMC_distance}). The enhancement of orbital poles increases with LMC infall mass, and is expected to last for at least 1 Gyr, but further studies are needed to assess its longevity. 

We have further characterized the physical origin of the simulated clustering of orbital poles, finding it to result from three effects: 1) the non-inertial reference frame of the MW's disk as a result of the shift in phase space of the inner halo (COM displacement and reflex motion); 2) the impact on the kinematics of particles located in the the DM dynamical friction wake and collective response  induced by the LMC; and 3) the bias in the density of orbital poles caused by the selection of tracers planar distributions. Our main conclusions are as follows:

\begin{enumerate}[leftmargin=*]

    \item \textbf{The combination of both the COM displacement and the reflex motion enhances the density of orbital poles non-homogeneously along a great circle and is stronger than either effect alone.}  
    The amplitude of the enhancement varies along a sinusoidal pattern when projected on the sky, reaching its maximum density close to the LMC's orbital pole. Shifts in the COM position and velocity that mimic the average effect of the LMC, result in clustering of orbital poles near the LMC's pole by $\delta \rho_{iso}\approx15\%$ from 50-150 kpc. At distances beyond 150 kpc the enhancement of orbital poles reaches $\delta \rho_{iso}\approx40\%$ relative to the isolated MW simulation (Figure~\ref{fig:poles_summary}).

    \item {\bf The distribution of orbital poles computed using particles in the southern and northern hemispheres are different, indicating the importance of the DM dynamical friction wake and collective response .} The distribution of orbital poles for particles selected in the northern hemisphere are well described by sinusoidal patterns that arise from our tests of the reflex motion and COM displacement. However, particles in the south have a completely different pattern owing to the presence of the DM dynamical friction wake and the Southern collective response . 
    The orbital poles of dynamical friction wake particles cluster near that of the LMC and VPOS.
    The density of poles is enhanced by a factor of $\delta \rho_{max}\approx$ 30\% with respect to the underdense regions (or $\delta \rho_{iso}\approx$15\% relative to particles chosen in the same volume in the isolated MW)
    This exercise cautions against the use of only corrections for reflex motion (without spatial corrections or accounting for the DM dynamical friction wake) to adjust observations for the impact of the LMC. 

    \item \textbf{Clustering of orbital poles on the sky can also be caused by a selection bias.} Orbital poles of objects orbiting the MW that are selected within a spatially planar configuration will be biased such that the distribution of their orbital poles will cluster at the projected location of the normal to the plane, and 180$^{\circ}$ from that plane. Our tests indicate that the inclusion of the LMC will augment this effect for particles in the $y-z$ Galactocentric plane (on average by $\delta \rho_{iso} \approx$20\%). 
\end{enumerate}

Our results highlight multiple ways in which the passage of the LMC about the MW impacts the phase-space properties of objects in equilibrium in the MW halo. These perturbations are intimately linked with the trajectory of the LMC through the halo. To study the dynamical state of the MW in cosmological context, including the MW's plane of satellites, one must include an LMC analog, which can bring in satellites of its own \citep[e.g.][]{Samuel20, santos20b} and may also enhance any existing plane through the effects described in this paper. The rarity of MW--LMC analogs with such configurations can explain why planes of satellites like that about our MW are so elusive in cosmological simulations.

All outer halo tracers ($\geq30$ kpc) should be impacted by the reflex motion \citep{Erkal20} and COM displacement (G19), but a subset that pass through the southern hemisphere in the vicinity of the DM dynamical friction wake and collective response  may be more affected, like Fornax and Sculptor \citep{Patel20}. Testing this assertion for specific satellites will require integrating orbits in time-evolving MW--LMC potentials that capture the DM dynamical friction  wake and collective response .
   
This work highlights the importance of constrained simulations that model the dynamics of specific systems, like the MW--LMC, to properly understand and interpret kinematic measurements. Similar simulations for the M31 and Centaurus-A system could bring new insights into the nature of the planes of satellites problem for other galaxy systems.

\acknowledgements{We thank the referee for their constructive comments, which helped improve the quality of this paper. We also thank Marcel Pawlowski for discussions and comments provided for an earlier version of this paper. We are grateful for the comments provided by Jenna Samuel, Alex Riley, Isabel Santos-Santos, and Veronica Arias. NG-C and GB are supported by NSF CAREER award AST-1941096, NASA ATP award 17-ATP17-0006, and HST grant AR 15004. EP is supported by the Miller Institute for Basic Research in Science at UC Berkeley. This work was supported in part by World Premier International Research Center Initiative (WPI Initiative), MEXT, Japan. CL acknowledges funding from the European Research Council (ERC) under the European Union’s Horizon 2020 research and innovation program (grant agreement No. 852839). 

All simulations were run on El-Gato 
at UArizona's High Performance Computing center, which is supported by the National Science Foundation under Grant No. 1228509.}
\software{}
Healpy\citep{healpy}
Scipy\citep{scipy}
Astropy\citep{astropy:2018}
Numpy\citep{numpy}

\bibliography{references} 
\bibliographystyle{aasjournal}

\end{document}